\documentclass[sigconf]{acmart}

\makeatletter
\def\@ACM@checkaffil{
    \if@ACM@instpresent\else
    \ClassWarningNoLine{\@classname}{No institution present for an affiliation}%
    \fi
    \if@ACM@citypresent\else
    \ClassWarningNoLine{\@classname}{No city present for an affiliation}%
    \fi
    \if@ACM@countrypresent\else
        \ClassWarningNoLine{\@classname}{No country present for an affiliation}%
    \fi
}
\makeatother

\usepackage{verbatim}
\usepackage{amsmath, amsfonts}
\usepackage{graphicx}
\usepackage{textcomp}
\usepackage{color}
\usepackage{subfigure} 
\usepackage{breakurl}
\usepackage{multirow}
\usepackage{diagbox}
\usepackage{makecell}
\usepackage{booktabs}
\usepackage{enumitem}
\usepackage{bbding}
\usepackage{xcolor}

\usepackage{xspace}
\usepackage{cleveref}
\usepackage{tablefootnote}
\usepackage{colortbl} 
\usepackage{pifont}

\newcommand{\etal}{{\textit{et al. }}}

\definecolor{mygray}{gray}{.9}
\newcommand*\colourcheck[1]{%
  \expandafter\newcommand\csname #1check\endcsname{\textcolor{#1}{\ding{52}}}%
}
\newcommand*\colourcross[1]{%
  \expandafter\newcommand\csname #1cross\endcsname{\textcolor{#1}{\ding{56}}}%
}
\colourcheck{black}
\colourcheck{gray}
\colourcross{black}
\colourcross{gray}

\AtBeginDocument{%
  \providecommand\BibTeX{{%
    \normalfont B\kern-0.5em{\scshape i\kern-0.25em b}\kern-0.8em\TeX}}}

\copyrightyear{2025}
\acmYear{2025}
\setcopyright{acmlicensed}
\acmConference[CCS'25] {Proceedings of the 2025 ACM SIGSAC Conference on Computer and Communications Security}{October 13--17, 2025}{Taipei, Taiwan.}
\acmBooktitle{Proceedings of the 2025 ACM SIGSAC Conference on Computer and Communications Security (CCS'25), October 13--17, 2025, Taipei, Taiwan}
\acmISBN{979-8-4007-1525-9/2025/10}
\acmDOI{10.1145/3719027.3744888}
\begin{document}
\title{Off-Path TCP Exploits: PMTUD Breaks TCP Connection \\Isolation in IP Address Sharing Scenarios}


\author{Xuewei Feng}
\affiliation{%
  \institution{Tsinghua University}
  \city{Beijing}
  \country{China}
}
\email{fengxw06@126.com}

\author{Zhaoxi Li}
\affiliation{%
  \institution{Tsinghua University}
  \city{Beijing}
  \country{China}
}
\email{li-zx24@mails.tsinghua.edu.cn}

\author{Qi Li}
\affiliation{%
  \institution{Tsinghua University}
  \city{Beijing}
  \country{China}
}
\email{qli01@tsinghua.edu.cn}

\author{Ziqiang Wang}
\affiliation{%
  \institution{Southeast University}
  \city{Nanjing}
  \country{China}
}
\email{ziqiangwang@seu.edu.cn}

\author{Kun Sun}
\affiliation{%
  \institution{George Mason University}
  \city{Fairfax}
  \state{Virginia}
  \country{USA}
}
\email{ksun3@gmu.edu}

\author{Ke Xu}
\authornote{Corresponding author}
\affiliation{%
  \institution{Tsinghua University}
  \city{Beijing}
  \country{China}
}
\email{xuke@tsinghua.edu.cn}

\renewcommand{\shortauthors}{Xuewei Feng et al.}

\begin{abstract}
%
%
%
%
%
Path MTU Discovery (PMTUD) and IP address sharing are integral aspects of modern Internet infrastructure.
In this paper, we investigate the security vulnerabilities associated with PMTUD within the context of prevalent IP address sharing practices.
We reveal that PMTUD is inadequately designed to handle IP address sharing, creating vulnerabilities that attackers can exploit to perform off-path TCP hijacking attacks.
We demonstrate that by observing the path MTU value determined by a server for a public IP address (shared among multiple devices), an off-path attacker on the Internet, in collaboration with a malicious device, can infer the sequence numbers of TCP connections established by other legitimate devices sharing the same IP address. This vulnerability enables the attacker to perform off-path TCP hijacking attacks, significantly compromising the security of the affected TCP connections.
Our attack involves first identifying a target TCP connection originating from the shared IP address, followed by inferring the sequence numbers of the identified connection.
We thoroughly assess the impacts of our attack under various network configurations. Experimental results reveal that the attack can be executed within an average time of 220 seconds, achieving a success rate of 70\%.
Case studies, including SSH DoS, FTP traffic poisoning, and HTTP injection, highlight the threat it poses to various applications.
Additionally, we evaluate our attack across 50 real-world networks with IP address sharing—including public Wi-Fi, VPNs, and 5G—and find 38 vulnerable.
%
Finally, we responsibly disclose the vulnerabilities, receive recognition from organizations such as IETF, Linux, and Cisco, and propose our countermeasures.

\end{abstract}




\maketitle


\section{Introduction}
\label{sec:intro}

PMTUD, defined in RFC 1191 and RFC 8201, is a crucial mechanism on the Internet designed to determine the maximum packet size that can traverse a network path without requiring IP fragmentation~\cite{rfc1191,rfc1981,rfc8201}. PMTUD is widely implemented by modern operating systems (OSes) to optimize data transmission, enhance performance, and reduce the overhead associated with packet fragmentation~\cite{luckie2010measuring}. By ensuring packets are appropriately sized for the entire network path, PMTUD improves efficiency and reliability across various applications.

In this paper, we uncover a side channel vulnerability in the PMTUD mechanism stemming from its inability to ensure isolation of TCP connections in scenarios involving IP address sharing, which are commonly observed in various network practices, including NAT (Network Address Translation), VPNs (Virtual Private Networks), load balancing, CDNs (Content Delivery Networks), Anycast, and shared hosting. This side channel vulnerability can be exploited to facilitate off-path TCP hijacking attacks.

Specifically, according to the specifications, the PMTUD mechanism (e.g., when enabled on a server) operates on a per-IP address basis to discover and maintain the path MTU for a remote client's IP address destination. As a result, it does not ensure isolation on a per-connection basis, especially when multiple connections share the same IP address.
We demonstrate that by monitoring the path MTU value determined by the server for a public IP address shared among multiple devices, a malicious device (referred to as a puppet in our context) can collaborate with an off-path attacker on the Internet. This collaboration enables the off-path attacker to infer sequence numbers of TCP connections belonging to other legitimate devices sharing the same IP address. As a result, the attacker can inject out-of-band fake TCP packets into the victim's connection, thereby manipulating the traffic.

%



To construct our attack, we first identify the target TCP connection by inferring the ephemeral source port associated with the shared IP address.
We find that \textit{Port Preservation} and \textit{Per-Destination Sequential Allocation} are two commonly used methods for ephemeral source port allocation in IP address sharing scenarios\footnote{In the \textit{Port Preservation} method, the original source port generated by the device is preserved. In the \textit{Per-Destination Sequential Allocation} method, a random ephemeral port is initially assigned, followed by sequential assignments for subsequent connections to the same server; refer to \S\ref{sec:bak-nat} for details.}. We demonstrate that these two methods are vulnerable.
We identify a timing side channel in the \textit{Port Preservation} method. If the source port of a TCP connection from the attacker-controlled malicious device (i.e., the puppet) conflicts with an established victim connection, the gateway holding the shared public IP address will look up its ephemeral source port pool to select a new one, causing a time delay (or even a connection failure if no available port is found). This observable delay or failure reveals the presence of a victim TCP connection on that port.
Regarding the \textit{Per-Destination Sequential Allocation} method, the off-path attacker can impersonate the server and send forged TCP packets with guessed port numbers to the gateway holding the public IP address. If the guessed port is correct, the packet is routed to the puppet. By analyzing the packet, the puppet determines the baseline ephemeral source port value. Since the gateway assigns ephemeral ports sequentially, if a port beyond this baseline is already in use when the puppet establishes a new connection, a victim TCP connection on that port is identified.

\begin{figure}[h]
	\vspace{-1mm}
	\begin{center}
		  \includegraphics[width=0.46\textwidth]{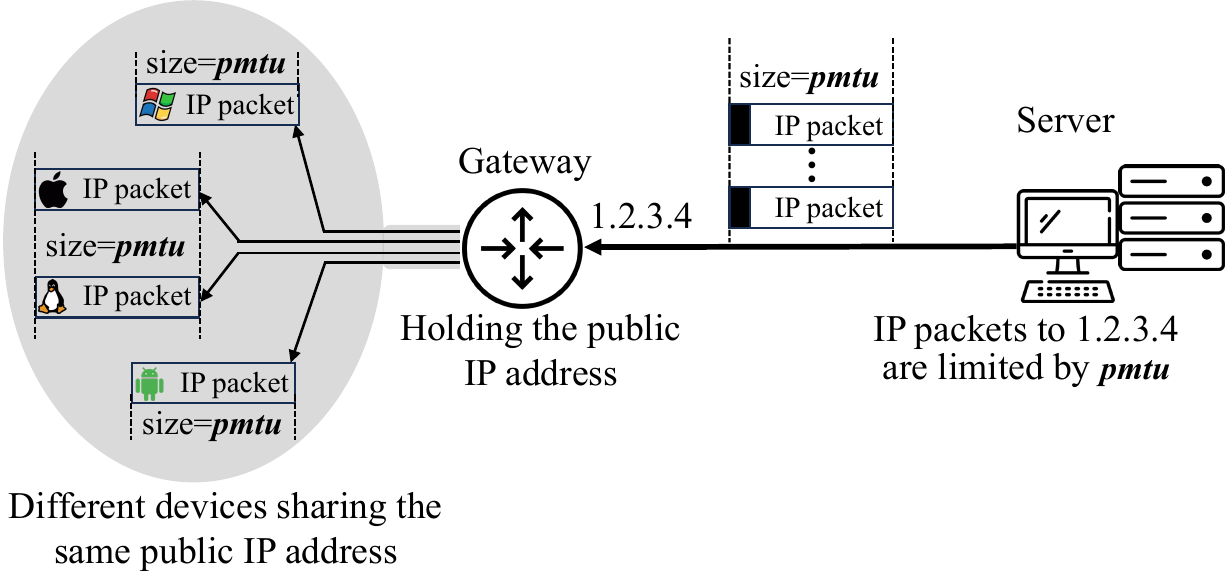}
		\vspace{-1mm}
		\caption{All devices sharing the same IP address are constrained by the server's path MTU for that address.}
		\label{pic:nat-pmtud}
	\end{center}
	\vspace{-2mm}
\end{figure}

Secondly, we proceed to infer the sequence numbers of the identified TCP connection to enable its off-path compromise.
%
In networks enabling IP address sharing (as shown in Figure~\ref{pic:nat-pmtud}), all devices share the same IP address and thus the same path MTU value established by the server for this IP address. This creates a side channel that can be exploited to infer the sequence numbers of TCP connections from victim devices.
Specifically, an off-path attacker on the Internet sends crafted ICMP error messages to the server. If the sequence number specified in the ICMP error message for the identified TCP connection is correct, the puppet observes a change in its own TCP packet length from the server. This occurs because the PMTUD mechanism operates on a per-IP address basis and cannot achieve per-connection isolation.
Once the source port and sequence number of the target TCP connection are identified, the attacker can inject fake TCP packets to compromise the connection. For example, the attacker could send crafted TCP \texttt{RST} packets to terminate the connection or poison the connection in scenarios where the acknowledgment number does not need to be precisely specified~\cite{Pan2024}, e.g., in the TCP implementations of operating systems such as macOS, Windows, and OpenBSD (refer to \S\ref{subsubsec:window-size} for detailed evaluations).

We conduct comprehensive evaluations of our attack, beginning with end-to-end tests to assess its effectiveness. Experimental results indicate that, with a puppet present in a targeted network where IP address sharing is enabled among multiple devices, an off-path attacker can identify a victim’s TCP connection and determine its sequence numbers with an average time of 194 seconds.
Additionally, we identify that victim devices running macOS, Windows, OpenBSD, and Linux have lenient checks on acknowledgment numbers, with an acceptable range larger than $2^{10}$. Therefore, attackers can send up to $2^{22}$ crafted TCP packets with different acknowledgment numbers simultaneously to evade the inference of the correct acknowledgment number~\cite{Pan2024}. If one packet's acknowledgment number is within the acceptable range, it will be accepted.
We show that with attack traffic bandwidth up to 23.95 Mbps, the attacker can terminate or poison the victim's connection using crafted TCP packets, achieving a success rate of 70\%.
Case studies, including SSH DoS, FTP traffic poisoning, and HTTP injection, demonstrate the threat our attack poses to real-world applications.
Additionally, we assess the impact on 50 real-world networks enabling IP address sharing, including NAT-enabled public Wi-Fi networks in coffee shops, hotels, and other locations, 5G cellular networks in different locations, as well as VPN Networks. Our measurements show that 38 of the 50 tested networks (over 75\%) are vulnerable.

Finally, we recommend countermeasures.
%
We reported our attack to the IETF, which reviewed our report and discussed potential mitigations with us. We recommend that PMTUD operate on a per-connection basis (instead of a per-IP address basis) to prevent information leakage between TCP connections.
In practice, different connections from a server may take different network paths to the same destination. Therefore, the path MTU of one connection should not limit all connections to that destination. Following IETF's suggestions, we will present our findings at an IETF working group meeting and propose standardization to address this vulnerability.
Additionally, we recommend that gateways implementing IP address sharing adopt a randomized method for allocating ephemeral source ports for TCP connection requests. This would prevent attackers from predicting port distributions, thereby thwarting attempts to identify victim TCP connections.
We have responsibly disclosed the vulnerable ephemeral source port allocation methods to the identified vendors, which have been recognized by Linux, Cisco, and H3C. In particular, Cisco issued a public statement (\url{https://bst.cisco.com/bugsearch/bug/CSCwm63019}) thanking us for enhancing the security of their devices.



%

\vspace{2mm}
\noindent \textbf{Contributions}. Our main contributions are the following:
\begin{itemize}[leftmargin=*]
\vspace{-2mm}
    
\item We reveal that PMTUD is inadequately designed to handle the prevalent practice of IP address sharing, leading to a fundamental side channel vulnerability that can be exploited to infer TCP sequence numbers.


\item We develop a novel off-path TCP hijacking attack that executes within an average time of 220 seconds, achieving a 70\% success rate and compromising critical applications such as SSH DoS, FTP traffic manipulation, and HTTP injection.

\item We responsibly disclosed the vulnerabilities, and propose countermeasures to mitigate the identified attack. 

\end{itemize}


\noindent \textbf{Ethics Statement:}
%
%
In our evaluations of the identified attack, we conducted a measurement study on 50 real-world networks that enable IP address sharing, such as public Wi-Fi networks and 5G cellular networks in various locations, to validate the effectiveness and impact of our attack. Ethical considerations were our top priority throughout the study. Specifically, we implemented several measures to ensure that our real-world tests did not impact other users or adversely affect the networks being tested.
First, when testing target networks that use public IP address sharing, we transparently communicated our experimental details to the network administrators and proceeded only after obtaining their approval.
Second, we ensured that all TCP connections under examination were exclusively ours, thus avoiding any interference with other connections. Our deployed devices established TCP connections with our VPS server, allowing us to perform inferences and analysis without disrupting regular users.
Finally, upon concluding our experiments, we reported our findings to the network administrators and reset our VPS server, restoring its path MTU value to the original setting for the public IP address used by the target network.

\section{Background}
\label{sec:background}

\subsection{Path MTU Discovery}\label{sec:bak-pmtud}

PMTUD is a networking mechanism used to dynamically determine the Maximum Transmission Unit (MTU) for a specific Internet  path from an IP originator to an IP destination~\cite{rfc1191,rfc1981}. The path MTU represents the largest packet size that can be transmitted from the originator to the destination without requiring IP fragmentation. PMTUD is essential for optimizing network performance by preventing packet fragmentation and reassembly, which can introduce inefficiencies and increase the risk of data loss.

\begin{figure}[h]
	\vspace{-1mm}
	\begin{center}
		\includegraphics[width=0.46\textwidth]{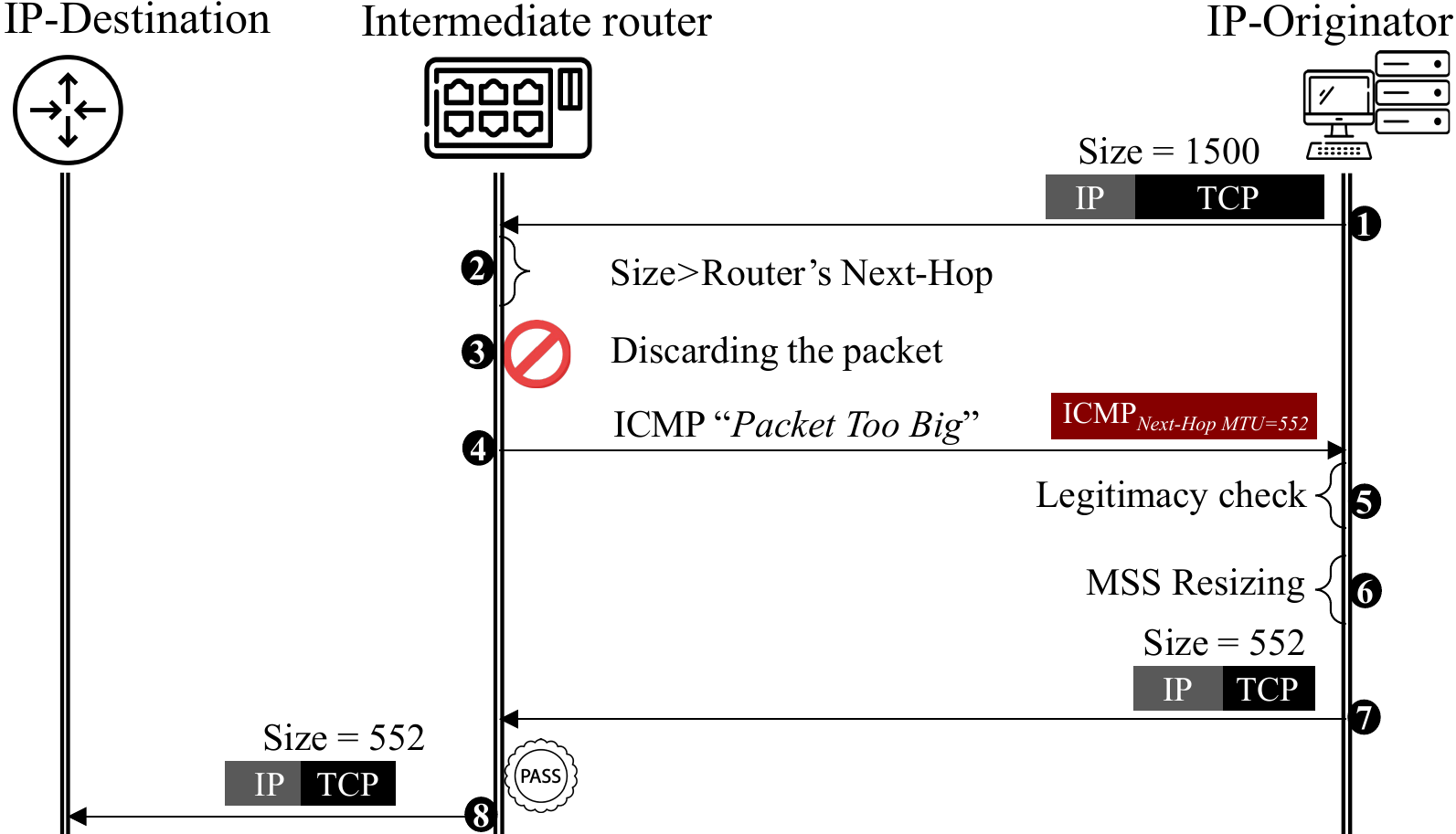}
		 \vspace{-2mm}
		\caption{Workflow of PMTUD to avoid IP fragmentation.}
		\label{pic:pmtud}
	\end{center}
	\vspace{-2mm}
\end{figure}

Figure~\ref{pic:pmtud} illustrates how PMTUD operates in the context of communication between an IP originator (e.g., a server on the Internet), an intermediate router, and an IP destination (e.g., a client or gateway holding a publicly routable IP address).
Initially, the originator sends TCP packets to the destination based on the previously established TCP connection. The size of the packets is limited to the originator's default next-hop MTU (e.g., 1500 octets), and the \texttt{DF} (Don't Fragment) flag in the packets' IP header is set to \texttt{True}.
The packet travels through several intermediate routers on its way to the destination. Each router along the path has its own next-hop MTU limit, which is the maximum packet size it can handle without fragmenting it. If the packet size exceeds the next-hop MTU of a router along the path, that router will discard the packet and issue an ICMP ``Packet Too Big'' message (ICMP error message with \texttt{Type}=3 and \texttt{Code}=4) to the originator.
The ICMP message will carry the discarded TCP packet, as well as the router's next-hop MTU value (e.g., 552 octets) in the low-order 16 bits of the ICMP header field labeled ``unused''.

\begin{figure}[h]
	\vspace{-2mm}
	\begin{center}
		\includegraphics[width=0.45\textwidth]{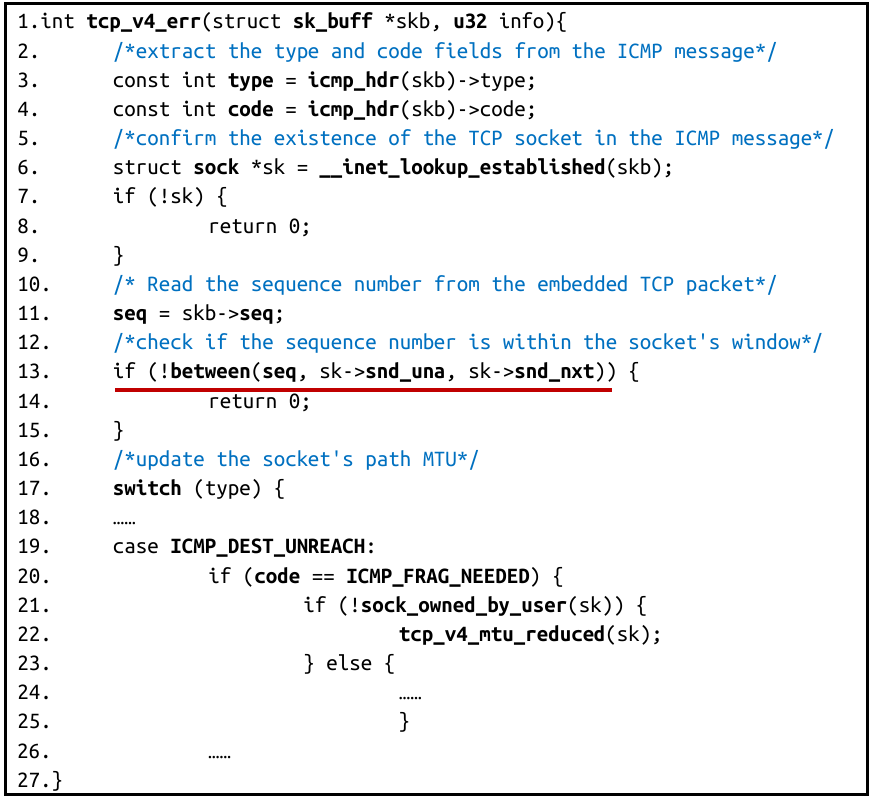}
		\vspace{-2mm}
		\caption{Verify the legitimacy of the received ICMP ``Packet Too Big'' message within Linux kernel 6.8.}
		\label{pic:code-snap}
	\end{center}
	\vspace{-2mm}
\end{figure}

According to the ICMP specifications \cite{rfc792,rfc1122,rfc1812,rfc5927}, upon receiving the ICMP ``Packet Too Big'' message, the originator will conduct a legitimacy check against the received message. This involves verifying at least 28 octets (i.e., 20 octets of the IP header plus at least the first 8 octets) of the original TCP packet that triggered the message.
Specifically, the originator verifies whether the sequence number of the TCP packet in the message falls within its sending window (e.g., in Linux systems, as shown in line 13 of Figure \ref{pic:code-snap}). Upon passing this legitimacy check, the originator updates the path MTU value stored at the IP layer for the destination, using the next-hop MTU value provided in the ICMP error message.
Meanwhile, the originator resizes its Maximum Segment Size (MSS) value based on the updated path MTU and reduces the size of subsequent TCP packets to the destination.
Finally, the router forwards the resized TCP packets to the next hop without IP fragmentation. The process of adjusting the MSS and resending packets continues until the originator determines the optimal packet size that can traverse the entire network path.
Once the originator successfully communicates with the destination using an MTU size that is compatible with the entire path, the PMTUD process concludes.

In summary, by listening to feedback from intermediate routers and adjusting packet sizes accordingly, PMTUD ensures efficient data transmission from an IP originator to an IP destination without causing IP fragmentation. However, in this paper, we reveal that PMTUD can be exploited by off-path attackers to infer the sequence numbers of TCP connections. In network scenarios where multiple devices share the same public IP address, any adjustments to the path MTU value made by the originator due to one device's TCP connection will affect the packet sizes of other devices' TCP connections. This alteration creates a side channel, potentially leaking confidential information from victim TCP connections.





\subsection{IP Address Sharing and Port Allocation}\label{sec:bak-nat}

IP address sharing is a widely adopted networking technique that addresses IPv4 address exhaustion and enables scalable, efficient Internet access for multiple devices and services. Technologies such as NAT, VPNs, load balancing, CDNs, and shared hosting rely heavily on IP address sharing. NAT allows home or business networks to connect multiple devices to the Internet using a single public IP address. This technique is also used in ISP mobile cellular networks to manage large numbers of users with limited IP resources. VPNs and proxy servers enable users to share a public IP, enhancing privacy and anonymity. In cloud services and large websites, load balancing distributes traffic across multiple servers using a single IP to ensure optimal performance. Similarly, CDNs use shared IPs to deliver content rapidly from geographically distributed servers. In shared hosting, multiple websites on the same server use one IP, differentiated by domain names, to provide cost-effective hosting solutions. These technologies are integral to various scenarios within Internet infrastructure.

\begin{figure}[h]
	\vspace{-3mm}
	\begin{center}
		  \includegraphics[width=0.44\textwidth]{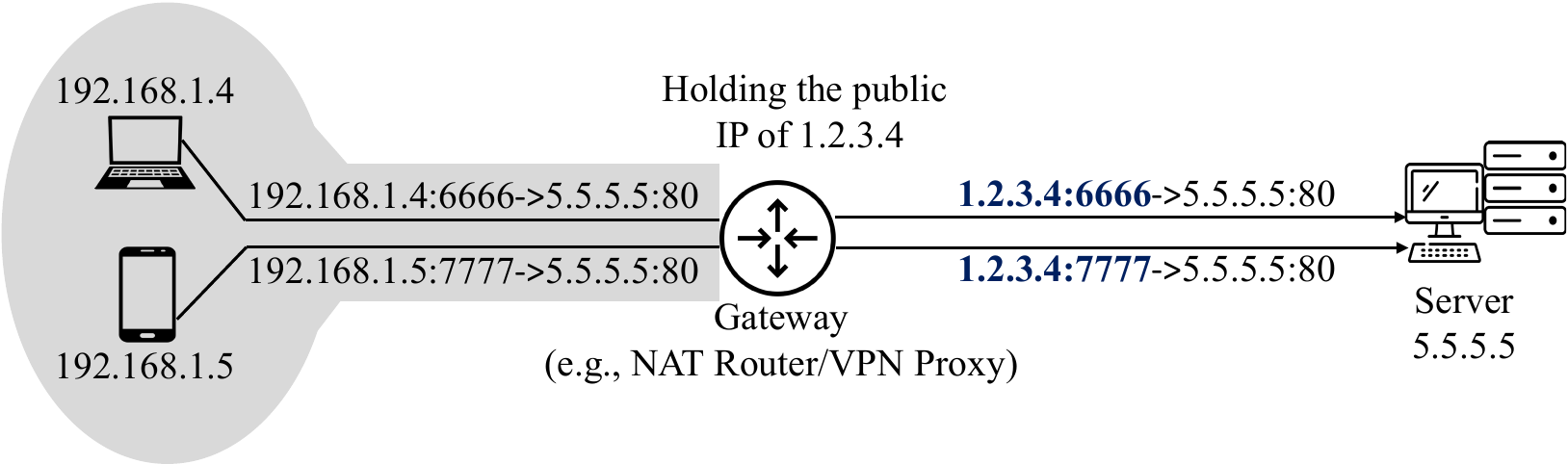}
		\vspace{-2mm}
		\caption{Multiple devices share the same public IP address to access a remote server on the Internet.}
		\label{pic:nat}
	\end{center}
	\vspace{-3mm}
\end{figure}

Figure~\ref{pic:nat} illustrates a typical scenario (e.g., a NATed network or a VPN network) in which multiple devices within a private realm share a single public IP address held by the gateway (i.e., a NAT router or a VPN proxy server) to access a server on the Internet.
When a device initiates a connection to the remote server, the gateway translates the device's private IP address and source port into its public IP address and a unique ephemeral port. This mapping is recorded in the gateway's session table to ensure that incoming responses from the server are correctly forwarded to the original device.

Particularly, ephemeral port allocation by the gateway significantly influences the behavior of IP address sharing, as it determines how the gateway rewrites the devices' original source ports and assigns new unique ephemeral source ports for outgoing TCP connection requests, allowing multiple devices to share a single public IP address. Three primary methods are commonly employed~\cite{mixon2024attacking,Herzberg2012SecurityOP,rfc4787,rfc5382}, as shown in Table~\ref{tab:port-allocation}.
%

\begin{table}[h]
\caption{Port allocation in IP address sharing scenarios.}
\vspace{-3mm}
\centering
\label{tab:port-allocation}
\resizebox{0.46\textwidth}{!}{%
\begin{tabular}{@{}cll@{}}
\toprule
\textbf{No.} & \multicolumn{1}{c}{\textbf{Port Allocation}} & \multicolumn{1}{c}{\textbf{Employed Systems}}             \\ \midrule
1            & \textit{Port Preservation}                            & Linux and the derivatives                                 \\ \midrule
2 & \textit{Per-Destination Sequential Allocation} & \begin{tabular}[c]{@{}l@{}}HUAWEI Versatile Routing Platform\\ H3C router firmware\\ Cisco IOS\end{tabular} \\ \midrule
3            & \textit{Random Allocation}                            & \begin{tabular}[c]{@{}l@{}}FreeBSD\\ OpenBSD\end{tabular} \\ \bottomrule
\end{tabular}%
}
\end{table}

\textit{Port Preservation} prioritizes maintaining the original source port of TCP connections, as illustrated in Figure~\ref{pic:nat}. However, conflicts can occur when multiple devices attempt to use the same source port to connect to the same server. In such cases, the gateway typically assigns another available port for the new connection. We observe that this method is predominantly employed by Linux-based gateways and their derivatives.

\textit{Per-Destination Sequential Allocation} assigns a random ephemeral port to the initial connection for each remote server, with subsequent connections to the same server being assigned sequentially incremented ports. Our tests show that this method is widely adopted by ISP routers (e.g., HUAWEI Versatile Routing Platform, Cisco IOS, and H3C router firmware) and is thus common in ISP networks that enable IP address sharing (e.g., CGNAT-enabled 5G cellular networks), where many devices may be accessing the gateway to connect to the same remote server simultaneously. This approach effectively avoids performance loss due to port conflicts.
%

Conversely, \textit{Random Allocation} indiscriminately translates original source ports to randomly selected available ports. We test that FreeBSD and OpenBSD systems commonly use this method.

We observe that the \textit{Port Preservation} and \textit{Per-Destination Sequential Allocation} methods are widely adopted by real-world gateways that enable IP address sharing.
This is particularly evident since many gateway implementations are based on Linux systems\footnote{A large number of VPN proxy servers (e.g., the popular WireGuard, PPTP, L2TP, and OpenVPN) and load balancing systems that enable IP address sharing are developed based on Linux systems. Moreover, Linux is widely used in embedded firmware enabling IP address sharing. For instance, OpenWrt, a Linux-based operating system designed for embedded devices, has spawned more than 20 derivative projects and provides core firmware for over 2,000 types of NAT devices that enable IP address sharing (\url{https://openwrt.org/toh/start}).}.
Besides, ISP networks that enable IP address sharing and use ephemeral port sequential allocation provide convenient Internet access. In this paper, we uncover that TCP ephemeral source ports using these methods in such networks can be easily inferred.

\section{Threat Model}
\label{sec:overview}

\begin{figure}[h]
	\vspace{-2mm}
	\begin{center}
            \includegraphics[width=0.44\textwidth]{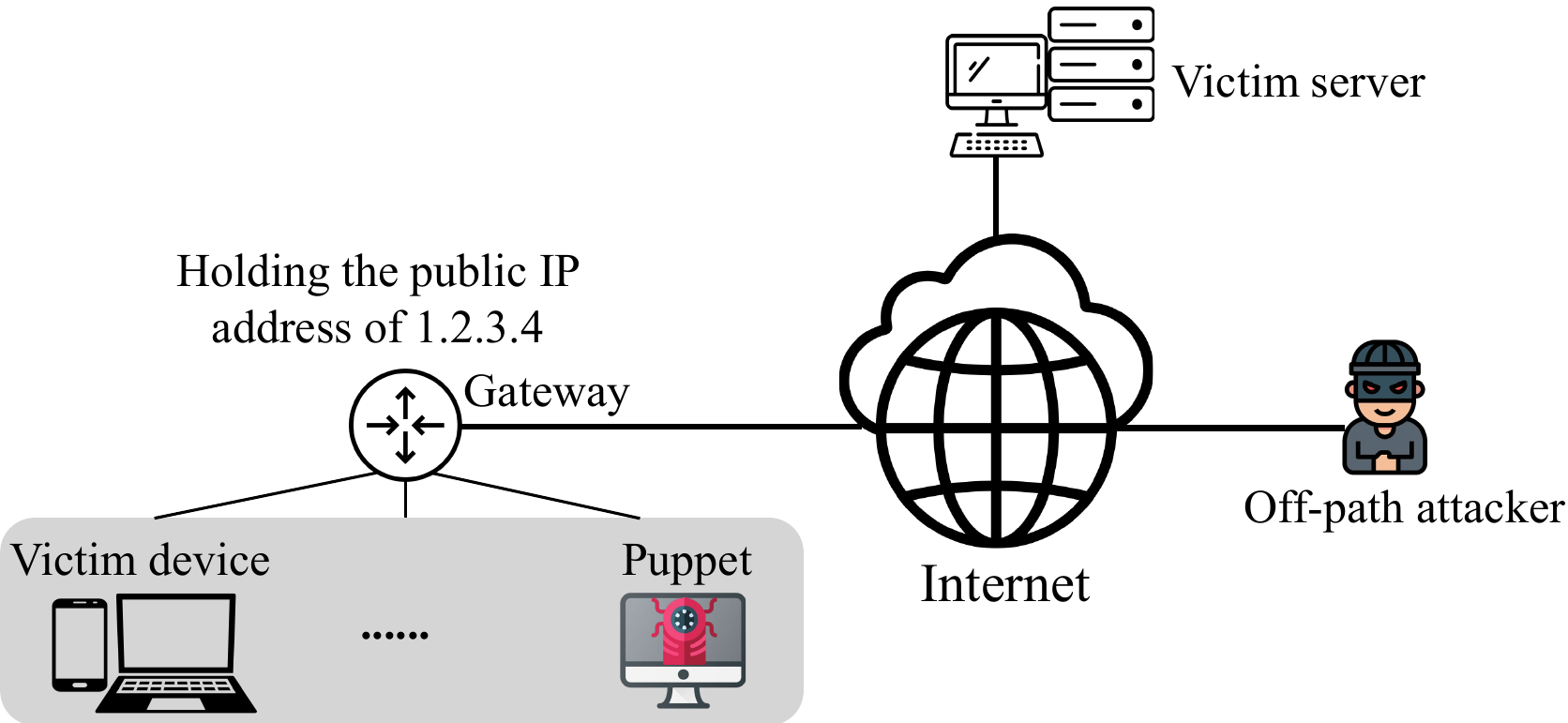}
		\vspace{-1mm}
		\caption{Threat model of hijacking TCP via side channels in the PMTUD mechanism.}
		\label{pic:threat-model}
	\end{center}
	\vspace{-2mm}
\end{figure}

As shown in Figure~\ref{pic:threat-model}, our attack involves five types of devices, i.e., a \textit{victim server}, a \textit{gateway} holding the shared public IP address, a \textit{victim device}, an \textit{off-path attacker}, and a \textit{puppet} (i.e., a malicious device controlled by the attacker within the IP address sharing realm).

The \textit{victim server} on the Internet provides various TCP-based services, such as SSH, web, and FTP. In this paper, we assume the server's IP address is 5.5.5.5, with known open TCP ports for requests, such as port 22 for SSH, port 20 for FTP downloads, or port 80 for the web.

The \textit{gateway} holds the public IP address (e.g., 1.2.3.4 in our example) and is responsible for sharing this IP address among multiple devices connected to it. It assigns a unique ephemeral port to each outgoing TCP connection, ensuring that the TCP connections from different devices using the same public IP address can be distinguished and maintain seamless communication with the victim server. In practice, the gateway may be a NAT-enabled router, a VPN proxy server, or a load balancing scheduler that facilitates public IP address sharing among multiple devices.


The \textit{victim device}, such as a mobile phone or a laptop connected to a NAT-enabled network or a VPN proxy server, is located behind the gateway. The victim device accesses services on the victim server through a TCP connection, and the source IP address and source port of the connection are rewritten by the gateway to the shared publicly routable IP address and a unique ephemeral port.

The \textit{puppet}, a malicious device controlled by the off-path attacker, connects to the same gateway as the victim device and shares the same public IP address.
In practice, the puppet could be implemented as a cellphone deployed by the attacker within a target NAT-enabled public Wi-Fi network or 5G cellular network, or as a malicious VPN client connected to a VPN proxy server.
The puppet establishes its own TCP connections with the victim server. The puppet-based threat model is widely used in off-path attacks~\cite{Herzberg2012SecurityOP,rytilahti2020using,chen2018off,feng2022ndss}.


%
The \textit{off-path attacker} on the Internet, with the assistance of the puppet, aims to identify the TCP connection between the victim device and the victim server and then infer the sequence numbers of the connection.
Subsequently, the attacker can terminate the victim's TCP connection using crafted TCP \texttt{RST} packets or inject crafted TCP data packets into the connection to manipulate the traffic. The attacker is capable of IP address spoofing~\cite{luckie2019network,cao2016off,man2021dns}.
%

%

It is worth noting that our attack differs from prior LAN (Local Area Network) attacks, such as ARP poisoning~\cite{sotillo2006ipv6}, malicious ICMP redirects~\cite{feng2023man}, TCP hijacking in Wi-Fi networks~\cite{yangexploiting}, or port exhaustion leading to DoS in NAT networks~\cite{nguyen2018slow}, which can be prevented by existing security measures (e.g., MAC-IP bindings~\cite{rahman2014holistic}, AP isolation~\cite{AP-isolation}, and reverse path validation~\cite{rfc2827,rfc3704}).
These security measures are less effective against our attack, as our attack does not require the puppet (which may not be located on the same LAN as the victim device, e.g., a malicious VPN client distributed remotely) to interact directly with the victim device or rely on IP address spoofing. Instead, the puppet only needs to maintain its own TCP connections with the victim server.

In practice, a TCP connection is identified by a four-tuple: [source IP address, source port number, destination IP address, destination port number]. In the off-path threat model, the destination IP address (i.e., the victim server) and destination port number are usually publicly known (e.g., port 80 of a popular HTTP server or port 22 of an SSH server)~\cite{cao2016off,chen2018off,cao2018off,ccsfeng,feng2021off}.
Moreover, in network scenarios involving IP address sharing, the source IP address is the public IP address held by the gateway, which can be easily obtained with the assistance of the puppet. Therefore, for the attacker, the ephemeral source port number in the four-tuple is the only value that needs to be inferred to identify a victim TCP connection.
Once a victim TCP connection is identified, the attacker also needs to infer the random sequence and acknowledgment numbers to inject a TCP packet into the connection, as modern TCP implementations typically use randomization mechanisms to mitigate blind out-of-band injection attacks.
However, we observe that weak checks on acknowledgment numbers are enforced in the TCP implementations of popular operating systems, allowing the attacker to bypass the need to infer acknowledgment numbers (refer to \S\ref{subsubsec:window-size} for detailed evaluations), as also evidenced by existing work~\cite{Pan2024}.
Consequently, in our attack, the attacker first infers a victim TCP connection by determining the ephemeral source port in the TCP four-tuple and then infers the sequence numbers to craft a disruptive TCP packet. In the next two sections, we elaborate on these two steps and show that vulnerabilities arising from interactions between PMTUD and IP address sharing can assist the attacker in achieving these objectives.

\section{TCP Connections Inference}
\label{sec:sourceport}

In this section, we propose methods to enable the off-path attacker to infer the ephemeral source port number assigned by the gateway, using two common allocation methods: \textit{Port Preservation} and \textit{Per-Destination Sequential Allocation}, for the victim device's TCP connections.


\subsection{Inferring Ports in Preserved Allocation}

\begin{figure}[h]
	\vspace{-2mm}
	\begin{center}
            \includegraphics[width=0.475\textwidth]{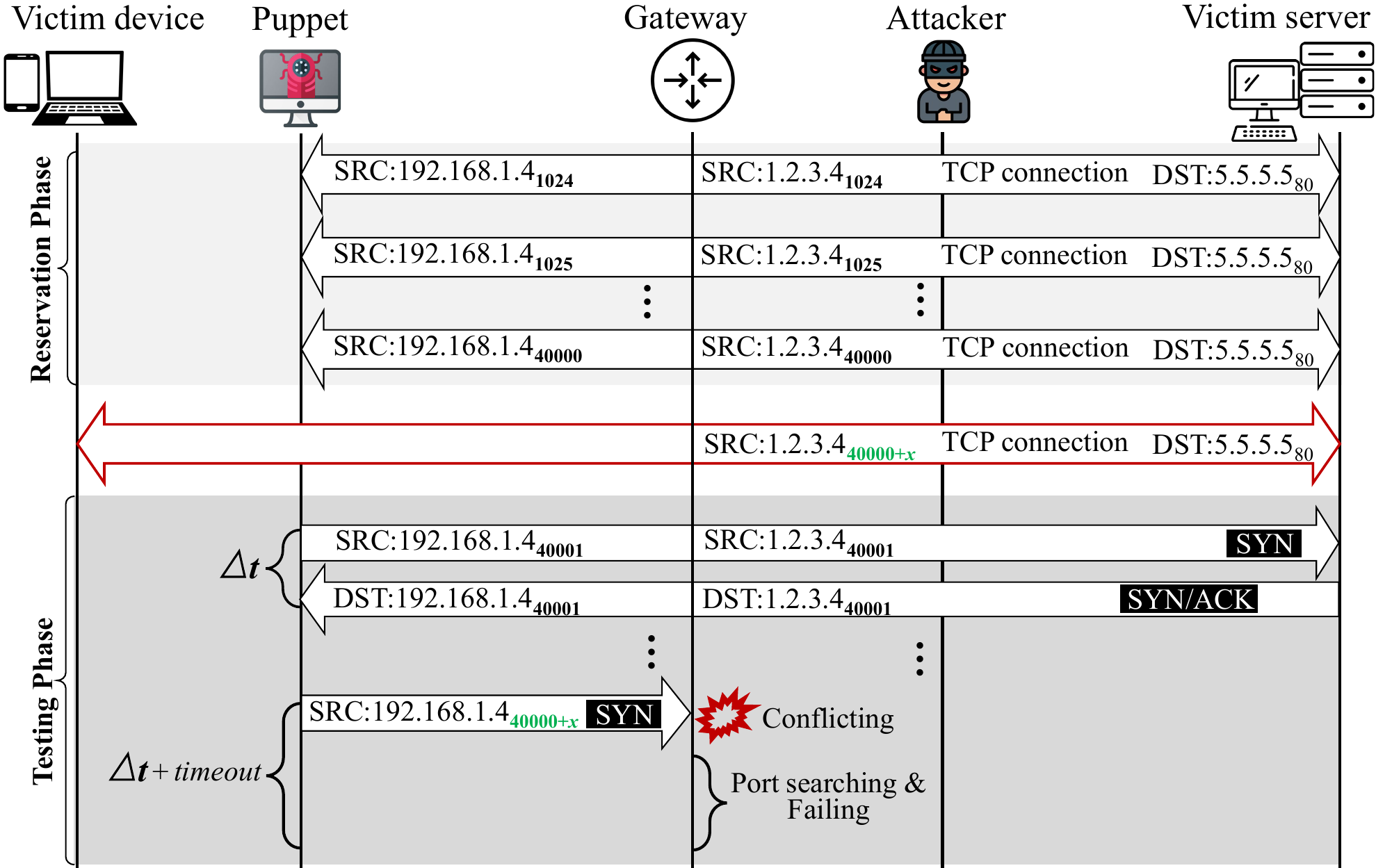}
		\vspace{-5mm}
		\caption{Identify the victim TCP connection on port preservation gateway.}
		\label{pic:Linuxport}
	\end{center}
	\vspace{-3mm}
\end{figure}

The \textit{Port Preservation} method, widely used by Linux-based gateways (e.g., OpenWrt routers and ExpressVPN servers), allocates ephemeral source ports for outgoing TCP connections by preserving the original source ports. We discover a timing side channel in this method that can be exploited to identify a victim's TCP connection.
As shown in Figure~\ref{pic:Linuxport}, our exploit consists of two phases: the \textit{Reservation Phase} and the \textit{Testing Phase}. 
At the beginning, the attacker deploys the puppet within the target IP address-sharing network. In the \textit{Reservation Phase}, the puppet initiates multiple TCP connections to the server, occupying a continuous range of source ports on the gateway (e.g., from 1024 to 40,000 \footnote{Ports below 1024 are reserved by Linux-based systems and are not used for ephemeral connections.}).
The gateway inserts entries into its session table to preserve the source ports of the puppet's TCP connections and manage these connections.
After this phase, the attacker and the puppet wait for a period for a victim device to establish its TCP connection with the server.
Since the puppet has occupied the source ports on the gateway from 1024 to 40,000, the gateway will allocate an ephemeral source port beyond this range (e.g., 40,000+$x$) to the victim connection.

In the \textit{Testing Phase}, the attacker sends TCP \texttt{SYN} packets with source ports one by one outside the reservation phase range, targeting the server's open port. Given Linux's method of reserving source ports, the gateway attempts to preserve the \texttt{SYN} packets' original source port as the ephemeral source port. If the source port (e.g., 40,001) does not conflict with the victim TCP connection's ephemeral source port (e.g., 40,000+$x$ and $x$\textgreater 1), the source port will be preserved, and the TCP \texttt{SYN/ACK} packet from the server will be quickly received by the puppet, allowing the puppet to test the time delay as $\Delta t$.
However, during the testing phase, if one of the TCP \texttt{SYN}'s source ports (i.e., 40,000+$x$) conflicts with the established victim connection, the gateway will randomly look up an idle port and allocate it for this TCP \texttt{SYN} packet, which is handled by the Netfilter framework in Linux-based systems.

Before Linux kernel version 5.0, the random allocation process exhaustively tried all ports within the available range (i.e., from 1024 to 65,535). Unless all ports were occupied, a connection could always be successfully allocated. However, because the port allocation algorithm operated within a software interrupt, too many attempts could lead to soft lockups, impacting other router functions. To address this issue, the Linux kernel updated the random allocation algorithm in version 5.0 and beyond, implemented in the file \texttt{nf\_nat\_core.c}.
In the new algorithm, if two devices under a Linux gateway initiate connections to the same server's port, specifying the same source port, the Linux kernel calls the \texttt{nf\_nat\_l4proto\_get\_unique\_tuple} function to resolve the conflict. This function employs a window-halving (starting with a window size of 128) search algorithm to look up available ports within the entire source port range (i.e., 1024 to 65,535).

Once a collision occurs (for example, if the gateway has already allocated ports 40,000+$x$ to the victim device's TCP connection and the puppet also requests to establish a new connection on this port), the function will first randomly select a port from the entire ephemeral source port range. Then, it checks if there is an idle port among the next 64 (i.e., half of the current window size of 128) consecutive ports that can be assigned to this new connection. If there are no idle ports, the window is halved (i.e., the window size becomes 32), and the process is repeated: randomly select a port from the entire ephemeral source port range and then check if there is an idle port among the next 32 consecutive ports. If there is still no port available, the function halves the current window size and repeats the process until an idle port is identified or the window size becomes smaller than 8. Consequently, after searching up to 5 rounds, if an available port is still not found, the Linux kernel will discard the \texttt{SYN} packet of the new connection.

Due to the puppet's port reservation in the prior phase, most ports of the gateway are continuously occupied, significantly reducing the success rate of random port allocation at this moment.
This may even result in the new TCP connection's (i.e., the puppet's new initiated TCP connection on port 40,000+$x$ in our example) \texttt{SYN} packet being dropped due to the failure of more than 5 rounds of searches. The delay or loss of the TCP \texttt{SYN} will result in the corresponding \texttt{SYN/ACK} packet being delayed or not being received by the puppet.
From the puppet's perspective, after the \texttt{SYN} packet is sent, the time for receiving the corresponding \texttt{SYN/ACK} packet is significantly delayed, i.e., greater than the previously tested $\Delta t$, and may even last until the corresponding entry in the gateway's session table for the \texttt{SYN} packet times out (typically 10 seconds in Linux systems). This delay allows the attacker to infer that the specified source port has been occupied by a victim TCP connection.

\subsection{Inferring Ports in Sequential Allocation}
\begin{figure}[h]
	\vspace{-1mm}
	\begin{center}
            \includegraphics[width=0.475\textwidth]{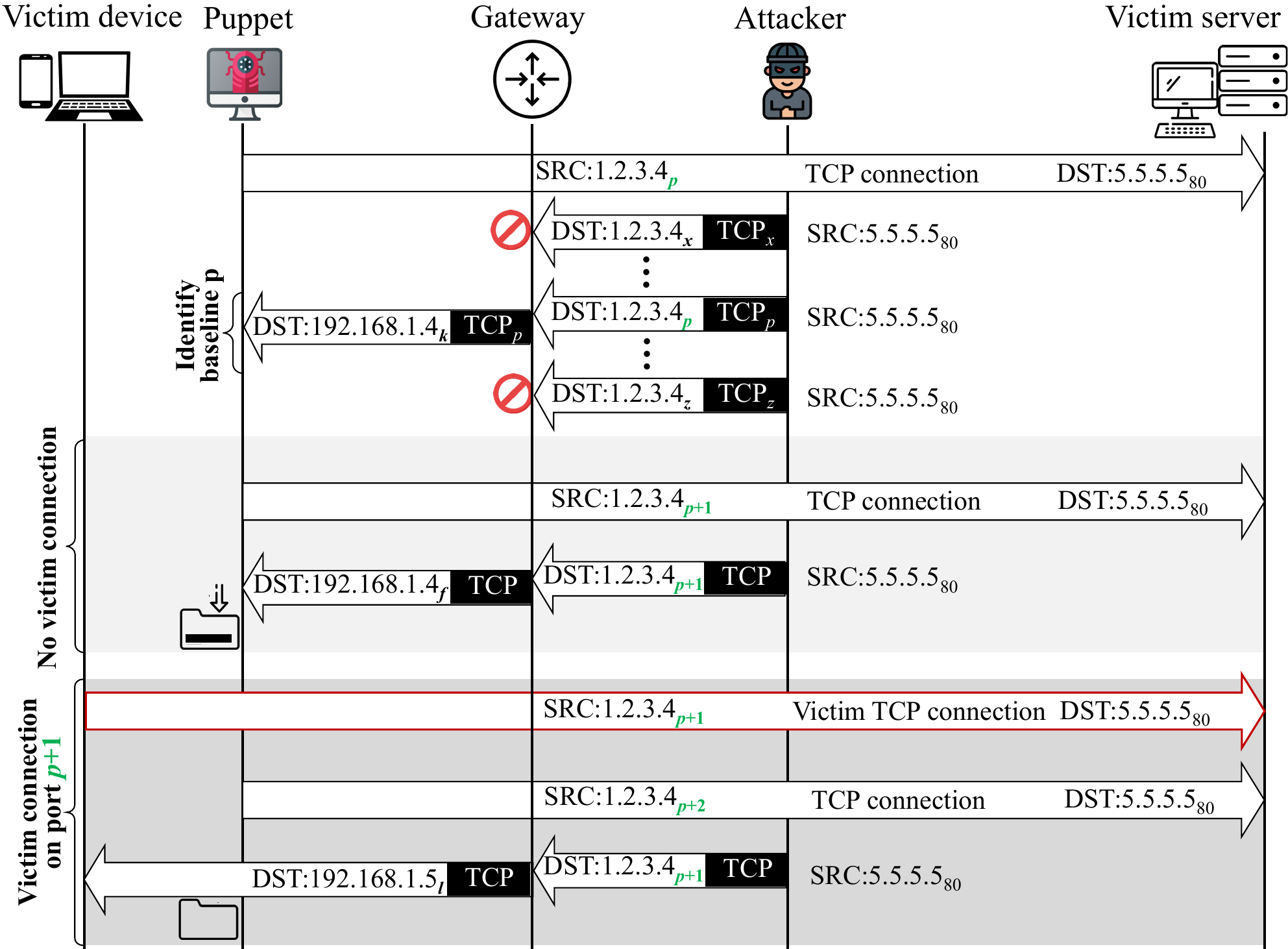}
		\vspace{-5mm}
		\caption{Identify victim TCP connection on port sequential allocation gateway.}
		\label{pic:portidentify}
	\end{center}
	\vspace{-2mm}
\end{figure}

Regarding the other widely used method, \textit{Per-Destination Sequential Allocation}, adopted by modern gateways for IP address sharing and ephemeral port allocation, Figure~\ref{pic:portidentify} illustrates our approach for identifying a victim TCP connection on a specific port of the gateway.
Initially, the puppet establishes a TCP connection with the remote server on the Internet (e.g., on port 80 with the IP address 5.5.5.5). Assuming the gateway (with the public IP address of 1.2.3.4) assigns an ephemeral source port number, denoted as $p$, to this TCP connection, this $p$ remains unknown to both the off-path attacker and the puppet.
After the connection is established, the attacker impersonates the server (via source IP address spoofing) and sends crafted TCP packets to the gateway's public IP address of 1.2.3.4. The destination port of each crafted TCP packet falls within the range of random ephemeral ports usually selected by TCP (e.g., 1024 $\sim$ 65,535).
Notably, the data section of each crafted TCP packet includes the destination port information of that packet. For instance, if the attacker designates port $x$ as the destination port for the current crafted TCP packet, it will embed the information ``TCP$_{x}$'' in the packet's data section. This information will assist the attacker in deducing the ephemeral source port assigned by the gateway for TCP connections to the server.


%


If the attacker crafts a TCP packet with the correct destination port (i.e., port $p$), the packet will ultimately reach the puppet after translation and routing by the gateway. Despite the gateway rewriting the address information in the packet (e.g., changing the destination port number from $p$ to $k$ and the destination IP from 1.2.3.4 to 192.168.1.4), the puppet can still extract the ephemeral source port number $p$ assigned to this connection from the packet's data section.
By contrast, crafted TCP packets with incorrectly destination port numbers will either be discarded by the gateway or routed to other devices and then discarded.

Once the puppet identifies the ephemeral source port number and informs the attacker, the attacker can establish a baseline value for source ports assigned by the gateway for connections to the victim server.
Subsequently, after a certain time interval, the puppet initiates a new TCP connection to the server. The ephemeral source port assigned by the gateway for this new TCP connection will vary based on whether the victim device has attempted to establish a TCP connection with the victim server within the mentioned time interval.
According to the \textit{Per-Destination Sequential Allocation} method, if the victim device did not initiate any TCP connection with the victim server during this interval, the gateway allocates the ephemeral source port $p$+1 for the new TCP connection created by the puppet. Consequently, if the attacker impersonates the server and sends crafted TCP packets to the gateway port $p$+1, the puppet will receive the crafted packet.
%
%
Instead, if the victim device initiated a TCP connection with the victim server during this interval, the gateway would have already assigned the source port number $p$+1 for that victim TCP connection.
In such a case, when the puppet establishes a new TCP connection with the server, the gateway will assign the ephemeral source port number $p$+2 for the puppet's new TCP connection (instead of $p$+1).
Consequently, if the attacker impersonates the server and sends crafted TCP packets to gateway port $p$+1, the puppet cannot receive the crafted packet because it is routed to the victim device.
%

Essentially, the off-path attacker detects the presence of a victim TCP connection on port $p$+1 of the gateway (holding the shared public IP address) by sending crafted TCP packets to that port. If the puppet receives the packet, it indicates the absence of a victim TCP connection. Conversely, if the puppet does not receive the packet, it implies that the port is already occupied, allowing the attacker to infer a victim TCP connection on port $p$+1 of the gateway.

\section{Sequence Numbers Inference}
\label{sec:sequence}

\begin{figure}[h]
	\vspace{-1mm}
	\begin{center}
            \includegraphics[width=0.475\textwidth]{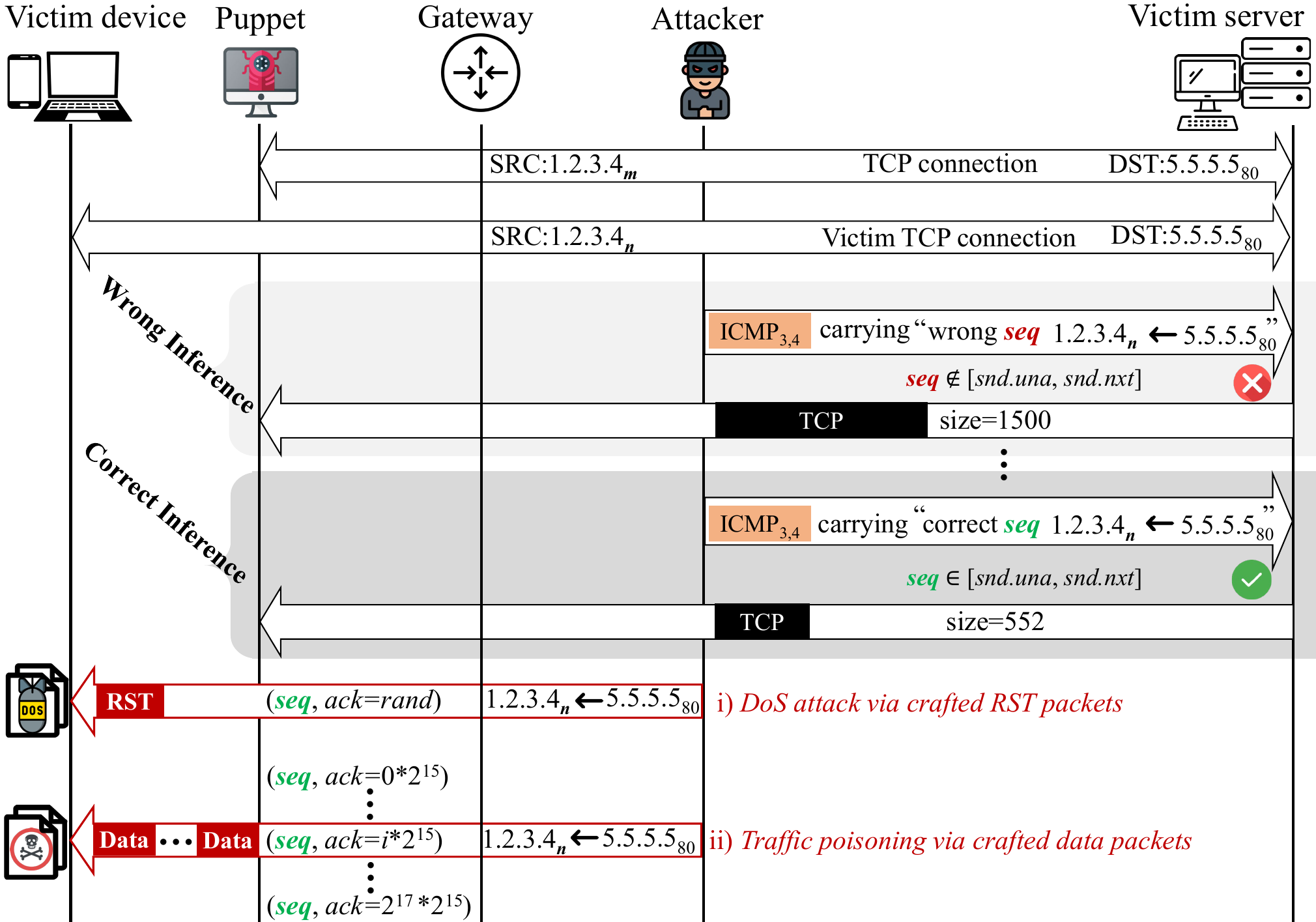}
		\vspace{-4mm}
		\caption{Infer sequence numbers of the victim TCP connection via the side channel in the PMTUD mechanism.}
		\label{pic:seqinfer}
	\end{center}
	\vspace{-1mm}
\end{figure}

In this section, we proceed to infer the sequence numbers of the identified TCP connection.
Figure~\ref{pic:seqinfer} illustrates the procedure of how to infer the sequence numbers of the victim connection. 
Initially, both the puppet and the victim device independently maintain their own TCP connections to the server, with the gateway assigning ephemeral source ports to these connections (e.g., $m$ and $n$, representing general cases including port preservation allocation method).
The off-path attacker's goal is to infer the sequence numbers of the identified victim TCP connection on port $n$. The attacker exploits a side channel tied to the path MTU value set by the server for the public IP address (e.g., 1.2.3.4), which is shared by both the puppet and the victim device.

\begin{figure}[h]
	\vspace{-2mm}
	\begin{center}
		  \includegraphics[width=0.46\textwidth]{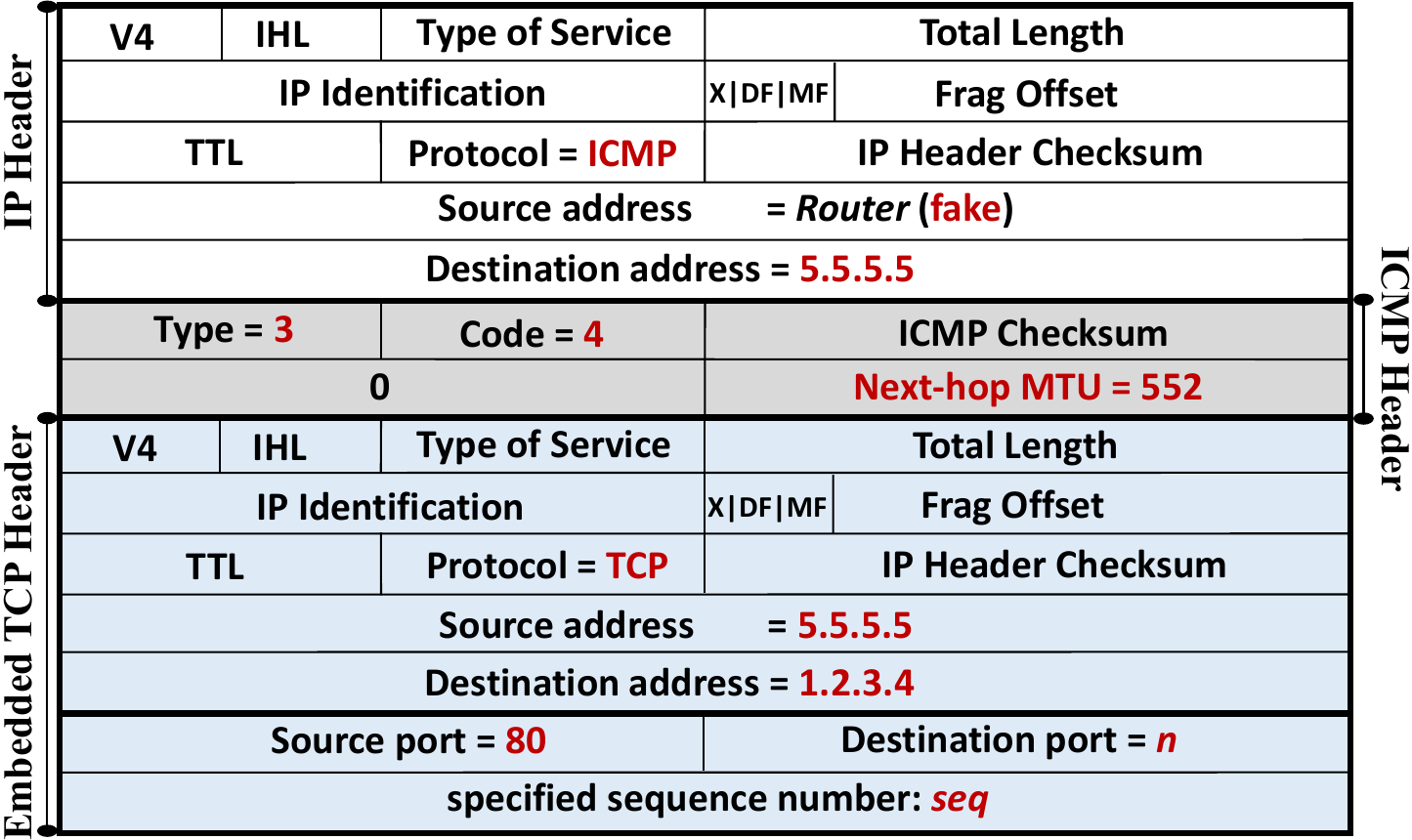}
		\vspace{-4mm}
		\caption{The crafted ICMP ``Packet Too Big'' message sent to the victim server by the attacker.}
		\label{pic:icmp}
	\end{center}
	\vspace{-2mm}
\end{figure}

At first, the attacker impersonates an intermediate router and sends a crafted ICMP ``Packet Too Big'' message (ICMP error message with \texttt{Type}=3 and \texttt{Code}=4), as shown in Figure~\ref{pic:icmp}, to the server. 
This message tricks the server into believing that the TCP packet it transmitted to the victim device was discarded by this intermediate router due to its excessive size. The crafted ICMP error message includes the inferred TCP packet header of the victim connection, which encompasses the server's sequence number, as well as IP addresses and port information.
Because ICMP error messages can be generated by any intermediate router along the transmission path~\cite{ccsfeng,feng2022off-redirect,brandt2018domain}, the attacker can craft these messages without resorting to IP spoofing. Additionally, in accordance with the ICMP specifications~\cite{rfc792,rfc1812}, ICMP error messages must carry at least the first 28 octets of the triggering packet, conveniently including the TCP port and sequence number information used by the server to associate the message with the appropriate process.

As shown in Figure~\ref{pic:seqinfer}, if the attacker specifies an incorrect sequence number in the TCP header embedded within the spoofed ICMP error message—i.e., a value outside the server’s sending window ([$snd.una$, $snd.nxt$], where $snd.una$ is the oldest unacknowledged sequence number and $snd.nxt$ is the next sequence number to be sent)—the server will discard the message as it fails the legitimacy checks of the PMTUD mechanism (see \S\ref{sec:bak-pmtud}).
Consequently, the server will not update the path MTU value associated with the gateway’s public IP address (e.g., 1.2.3.4) based on the MTU indicated in the ICMP message (e.g., ``Next-hop MTU=552'' in Figure~\ref{pic:icmp}). From the puppet's perspective, the size of TCP packets received from the server through its own connection remains unchanged (e.g., 1500 octets, as before).

In contrast, if the attacker specifies the correct sequence number in the ICMP message, that is, the sequence number happens to fall within the server's sending window ([$snd.una$, $snd.nxt$]) for the victim TCP connection, then the message will pass the legitimacy checks of the server's PMTUD mechanism. Furthermore, the server will update the path MTU value preserved for the corresponding IP address (i.e., the public IP address 1.2.3.4 of the gateway) based on this message, and adjust the size of all subsequent packets sent to that IP destination to not exceed this path MTU value.
From the puppet's perspective, it will observe a change in the size of the received TCP packets from the server (e.g., from the original 1500 octets decreased to the attacker-specified 552 octets). By observing this change, the attacker can identify that it have correctly deduced the sequence number of the victim TCP connection. 
%
Once the attacker identifies the sequence number of the victim TCP connection, it can terminate the connection by sending a forged \texttt{RST} packet with the guessed sequence number embedded, initiating a DoS attack.
According to the TCP specifications~\cite{rfc5961,rfc9293}, such \texttt{RST} packets only need to carry the correct sequence number and do not necessarily require an acknowledgment number.
Alternatively, the attacker can engage in traffic poisoning attacks by brute-forcing acknowledgment numbers to construct TCP data packets with the guessed sequence number embedded.

In practice, the attacker can speed up sequence number inference via an iterative probing strategy: partitioning the sequence number space, embedding selected values into spoofed ICMP messages, and sending them at a high rate (e.g., around 10,000 packets per second in our experiments) to the victim server. If the server responds with smaller TCP segments that reflect the reduced MTU, the attacker narrows the candidate sequence number range and further lowers the MTU value in subsequent probes. This process can eventually identify the server’s sending window and reveal the sequence number.

\section{End-to-End Evaluations}
\label{sec:evaluations}

In this section, we conduct a comprehensive end-to-end evaluation of our attacks. 
%
%
This involves evaluating the time and attack traffic bandwidth required to identify the ephemeral source port of a victim TCP connection, inferring the sequence numbers of the identified victim TCP connection, and examining the acceptable TCP acknowledgment numbers across different operating systems.

Moreover, we present three case studies (SSH DoS, FTP poisoning, and HTTP injection) to demonstrate the broad applicability and severity of our off-path attack against widely-used protocols. SSH and FTP remain prevalent in modern networks, with over 7 million active FTP servers currently deployed on the Internet according to SHODAN. We systematically analyze the success rates, time costs, and bandwidth costs associated with conducting these three attacks end-to-end.
Experimental results reveal that our attacks can be executed within an average time of 220 seconds, achieving a success rate of 70\%.

\subsection{Experimental Setup \& Workflow}

Our end-to-end experimental evaluations involve five devices: a victim device, a puppet, an attacker, a NAT gateway that allocates ephemeral ports and holds a shared public IP, and a victim server.
The puppet is equipped with Ubuntu 22.04 and Linux kernel version 5.15.
We configure the victim device with various OSes, including Linux, OpenBSD, macOS, and Windows (see \S\ref{subsubsec:window-size} for details) to test the range of acceptable acknowledgment numbers and evaluate the effectiveness of SSH DoS and FTP/HTTP injection attacks against them.
Both the puppet and the victim device are connected to the gateway.
%
%
The gateway employs one of two typical ephemeral source port allocation methods: \textit{Port Preservation} or \textit{Per-Destination Sequential Allocation}. For the \textit{Port Preservation} method, we use a gateway running OpenWrt 22.03 firmware, while for the \textit{Per-Destination Sequential Allocation} method, we use a gateway equipped with either HUAWEI USG6000, Cisco IOS 17.03.08 firmware, or H3C VSR1000 firmware. The gateway manages the translation and handling of external access requests from both the puppet and the victim device, including assigning the public IP address and allocating the ephemeral source port.
A VPS server running Ubuntu 22.04 and Linux kernel version 5.15 is in AS132203 of California, USA.
The puppet and victim device access services from the server, such as SSH (port 22), FTP (port 20), and web surfing (port 80).
%
%
An attack machine with Kali 2022.2 and source IP address spoofing capabilities attempts to identify the TCP connection between the victim device and server, infer the connection's sequence numbers, and hijack the victim's connection with the puppet's assistance.
%


In our end-to-end evaluations, we first infer the ephemeral source port assigned by the gateway to the victim device when it initiates a TCP connection to the server, and evaluate the time cost and attack traffic bandwidth needed for the attacker to infer this port under two allocation methods.
Subsequently, the attacker infers the sequence numbers of the identified victim TCP connection, and we analyze the time and bandwidth required for this step.
Once the correct sequence number is identified, the off-path attacker can send crafted TCP \texttt{RST} packets to launch an off-path DoS attack, terminating the victim TCP connection.
Next, we investigate the range of acceptable acknowledgment numbers for different OSes, providing the window sizes needed to craft TCP data packets to poison the target traffic.
Finally, we validate and test the end-to-end attack cost and success rate through three case studies: terminating SSH with crafted TCP \texttt{RST} packets carrying the inferred sequence number, and poisoning FTP/HTTP traffic with crafted TCP data packets carrying both the inferred sequence number and brute-forced acknowledgment number.
 
%
%
%

\subsection{Experimental Results}\label{subsec:experimental-results}

\subsubsection{Inferring Source Ports \& Identifying Connections}
We test the time and bandwidth required for the attacker to infer the ephemeral source port assigned by the gateway under two common port allocation methods, allowing identification of the victim TCP connection.

\begin{figure}[h]
        \vspace{-4mm}
	\begin{center}
        \subfigure[Time cost.]{
			\label{pic:linux_port_time}  
			\includegraphics[width=0.222\textwidth]{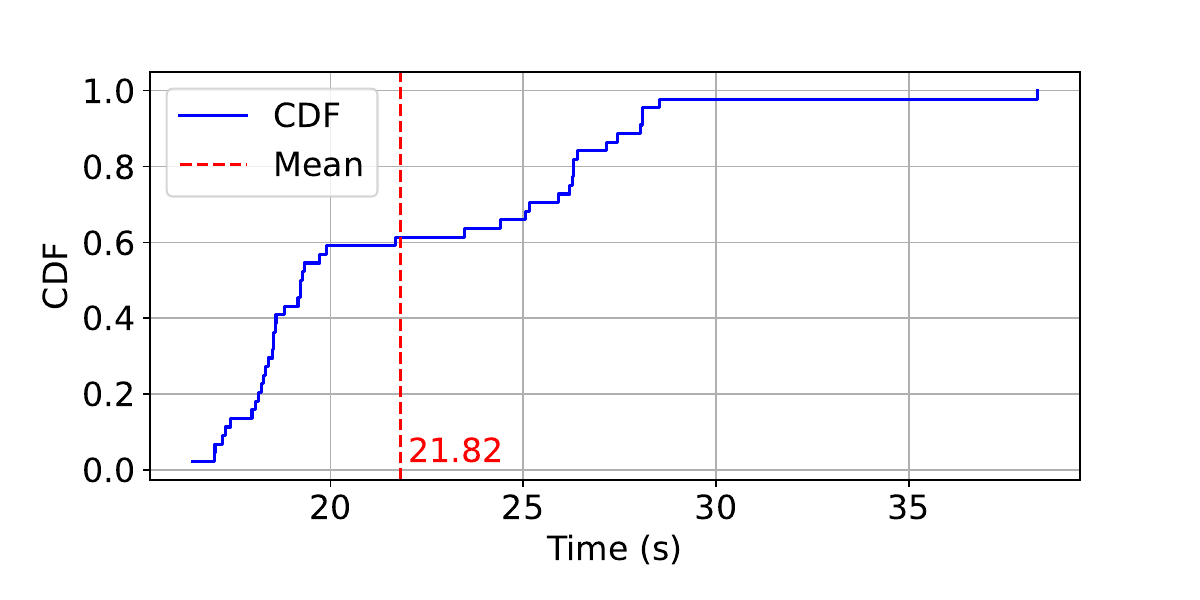} 
		} 
         \subfigure[Bandwidth cost.]{
			\label{pic:Linux_port_band}
			\includegraphics[width=0.222\textwidth]{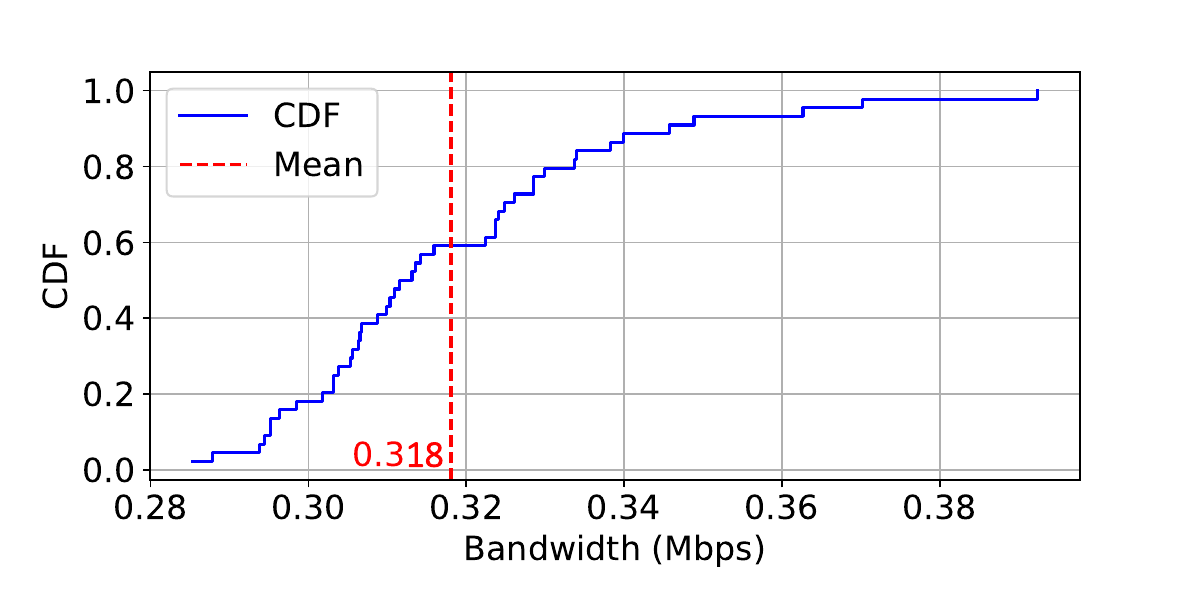}
		}
      \vspace{-4mm}
		\caption{CDF of attack cost for inferring the source port of victim TCP connections in the preserved allocation.} 
		\label{pic:Linux_port_cost} 
	\end{center}
	\vspace{-2mm}
\end{figure}

\textit{i) Cost for Inferring Source Port in Preserved Allocation.}
Figure~\ref{pic:Linux_port_cost} shows the CDF (Cumulative Distribution Function) of the time cost and attack traffic bandwidth required to identify the ephemeral source port assigned by the NAT gateway running OpenWrt 22.03 to the victim device's TCP connection.
We conduct 50 experiments, and the attacker, assisted by the puppet, typically requires an average of 21.82 seconds (Figure~\ref{pic:linux_port_time}) and 0.318 Mbps of bandwidth (Figure~\ref{pic:Linux_port_band}) to identify the ephemeral source port assigned by the gateway to the victim device's TCP connection.

\begin{figure}[h]
        \vspace{-4mm}
	\begin{center}
        \subfigure[Time cost.]{
			\label{pic:seq_port_time}  
			\includegraphics[width=0.222\textwidth]{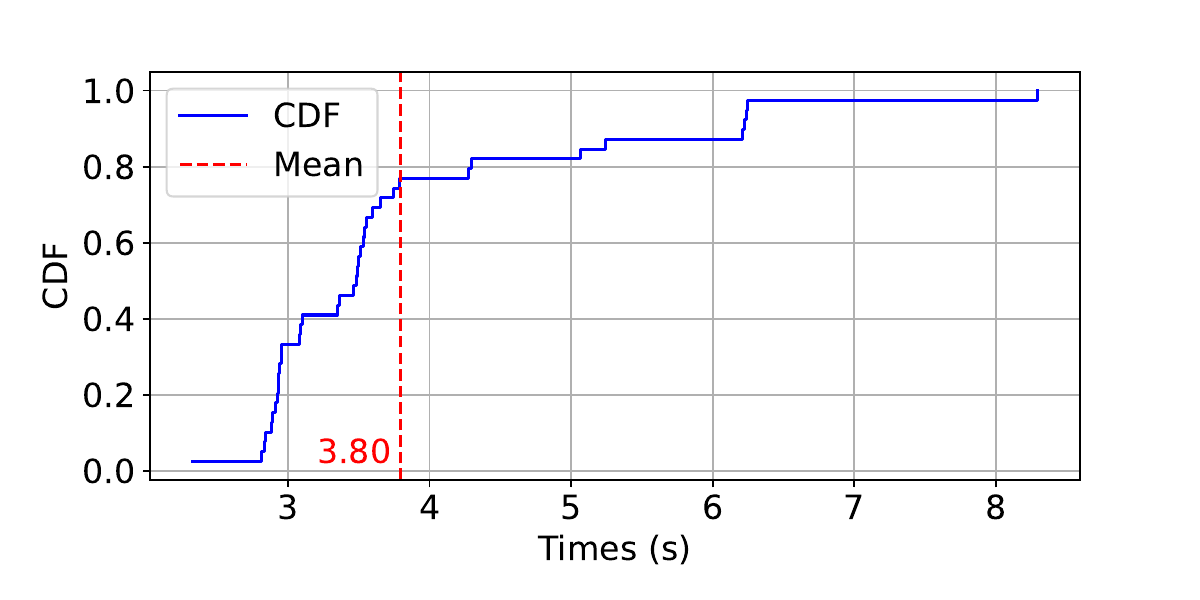} 
		} 
         \subfigure[Bandwidth cost.]{
			\label{pic:seq_port_band}
			\includegraphics[width=0.222\textwidth]{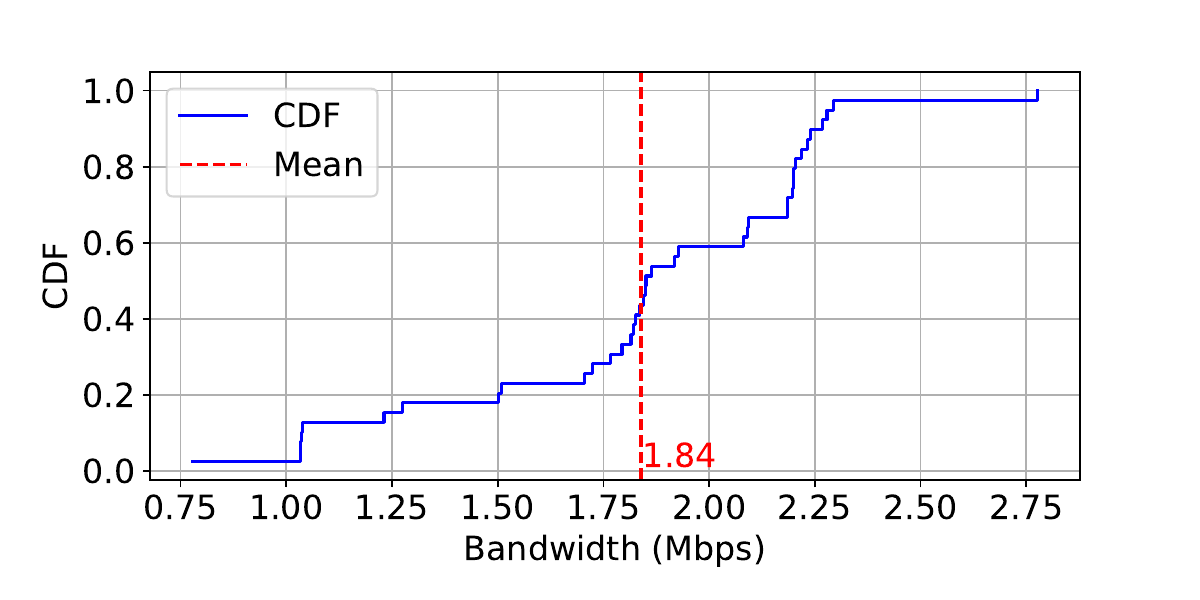}
		}
      \vspace{-4mm}
		\caption{CDF of attack cost for inferring the source port of victim TCP connections in the sequential allocation.} 
		\label{pic:seq_port_cost} 
	\end{center}
	\vspace{-4mm}
\end{figure}

\textit{ii) Cost for Inferring Source Port in Sequential Allocation.}
Figure~\ref{pic:seq_port_cost} shows the CDF of the time cost and attack traffic bandwidth needed to identify a target TCP connection initiated by the victim device connected to the gateway running Cisco IOS 17.03.08 or HUAWEI USG6000 firmware.
In our 50 experiments, the off-path attacker on the Internet, collaborating with the puppet, needs an average of 3.80 seconds (Figure~\ref{pic:seq_port_time}) and 1.84 Mbps of bandwidth (Figure~\ref{pic:seq_port_band}) to determine the baseline ephemeral source port assigned by the gateway for the victim device's TCP connections to the server. This enables the attacker to exploit the gateway's sequential port allocation method to identify the victim's TCP connection.

\subsubsection{Inferring Sequence Numbers}
\begin{figure}[h]
        \vspace{-4mm}
	\begin{center}
		\subfigure[Time cost.]{ 
			\label{pic:seq_time}  
			\includegraphics[width=0.222\textwidth]{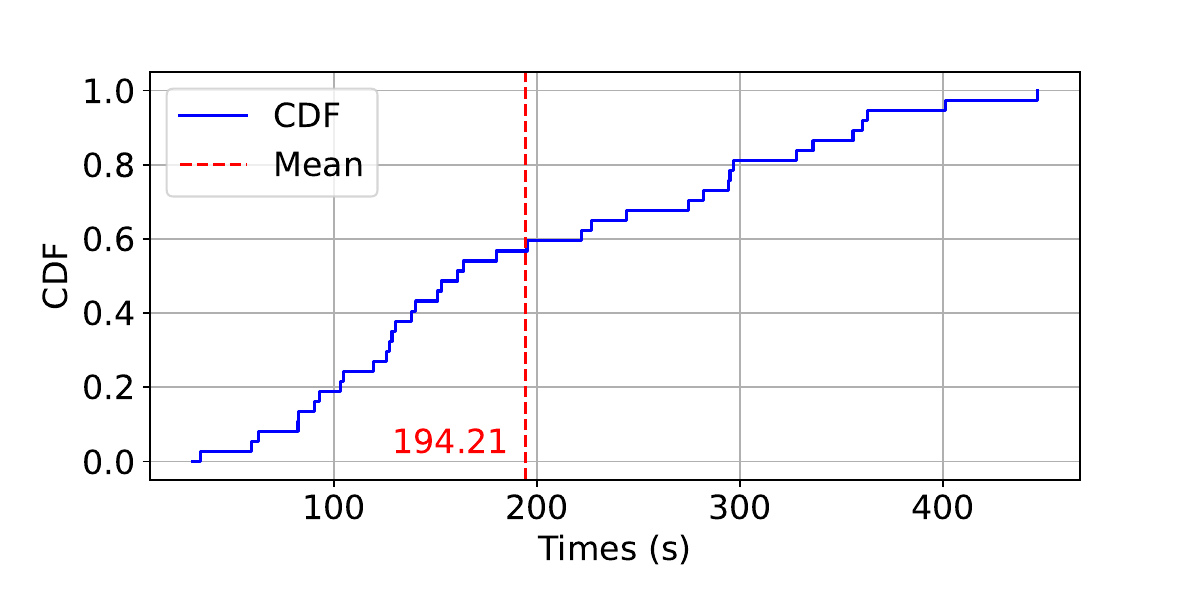} 
		} 
		\subfigure[Bandwidth cost.]{ 
			\label{pic:seq_band}
			\includegraphics[width=0.222\textwidth]{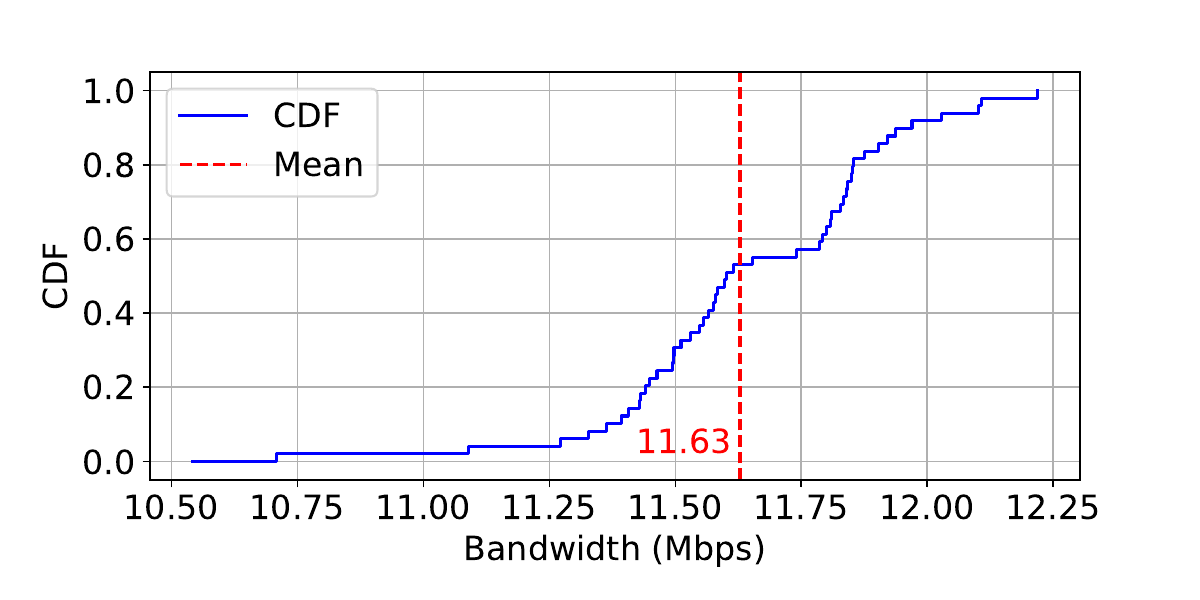}
		}
        \vspace{-4mm}
		\caption{CDF of time cost and bandwidth cost for inferring sequence numbers of the victim connection.} 
		\label{pic:seq_cost} 
	\end{center}
	\vspace{-3mm}
\end{figure}

After identifying the victim TCP connection, the attacker sends crafted ICMP ``Packet Too Big'' messages to the server, embedding a guessed sequence number of the victim TCP connection.
If the sequence number is correct, the path MTU value established by the server for the public IP address will be updated, i.e., decreasing from 1500 octets to 552 octets. This update affects the TCP packet size received by the puppet from the server. By observing these changes in packet sizes, the puppet and attacker can jointly infer the sequence number of the victim TCP connection.
Figure~\ref{pic:seq_cost} shows the CDF of time cost and attack traffic bandwidth for inferring the sequence number of the victim TCP connection. In our 50 experiments, the attacker needs an average of 194.21 seconds (Figure~\ref{pic:seq_time}) and 11.63 Mbps of bandwidth (Figure~\ref{pic:seq_band}) to identify the sequence number.

\subsubsection{Identifying the Range of Acceptable Acknowledgment Numbers}\label{subsubsec:window-size}
Once the sequence number of the victim TCP connection is identified, the attacker can send a crafted \texttt{RST} packet to terminate the victim TCP connection, constructing a DoS attack. However, for the attack to poison the target TCP traffic, the attacker must also guess the acceptable acknowledgment number of the target TCP connection to construct a valid TCP data packet. We investigate the acknowledgment number validation mechanisms of popular OSes and find that attackers can easily perform brute-force attacks on 32-bit acknowledgment numbers, mainly due to the lenient validation of ACKs allowed by current TCP implementations (such as the ``Ghost ACKs''~\cite{Pan2024}).

\begin{figure}[h]
	\vspace{-3mm}
	\begin{center}
            \includegraphics[width=0.22\textwidth]{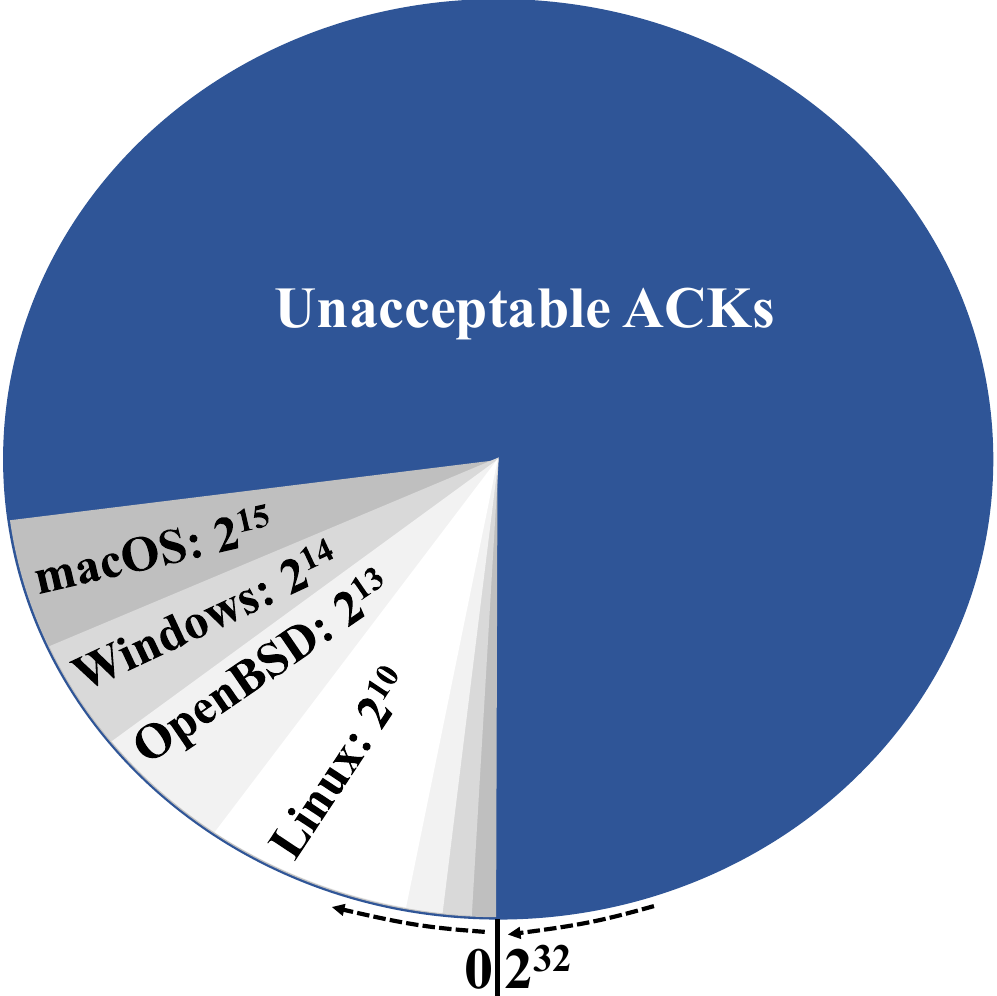}
		\vspace{-1mm}
		\caption{Acceptable acknowledgment number window sizes across different operating systems.}
		\label{pic:ackwindow}
	\end{center}
	\vspace{-3mm}
\end{figure}

Figure~\ref{pic:ackwindow} shows the window sizes of acceptable acknowledgment numbers for the different OSes we identified.
For macOS systems (macOS kernel version 14.2 in our test), the window size of the acceptable acknowledgment numbers is \(2^{15}\). Therefore, the attacker can divide the entire acknowledgment number space into \( \frac{2^{32}}{2^{15}} =  {2^{17}}\) windows and select one acknowledgment number within each window. This way, the attacker only needs to craft \({2^{17}} = 131,072\) TCP data packets, each carrying the guessed sequence number and one selected acknowledgment number, and send them to the victim TCP connection. One of these packets will be accepted by the victim client, ultimately resulting in traffic poisoning.

For Windows systems (Windows 11 Pro 23H2 in our test), the window size of the acceptable acknowledgment numbers is \(2^{14}\). For OpenBSD systems (kernel version 7.3), the window size is \(2^{13}\). For Linux systems (kernel version 5.15), the window size is \(2^{10}\).
By using this method, attackers can brute-force the acknowledgment number, eventually achieving a traffic poisoning attack. In the subsequent case studies of FTP poisoning and HTTP injection, we provide detailed information on the attack bandwidth and success rate for traffic poisoning against different OSes.

\subsection{Case Study}
%
We assess the end-to-end attack success rate through three case studies: terminating encrypted SSH connections with crafted TCP \texttt{RST} packets carrying only the inferred sequence number, and poisoning FTP and HTTP sessions with crafted TCP data packets carrying both the inferred sequence number and the brute-forced acknowledgment number.

\subsubsection{DoS against SSH}
According to TCP specifications~\cite{rfc5961,rfc793}, RST-based DoS attacks require the TCP \texttt{RST} packet to carry the exact sequence number (i.e., the next sequence number expected by the victim device, $rcv.nxt$).
We conduct 50 experiments, of which 38 are successful, with the attacker successfully terminating the identified SSH connection between the victim device and the server.
The failure of the attack is primarily due to the sliding of the server's sending window (e.g., caused by TCP keep-alive packets), which creates a discrepancy between the exact sequence number of the target connection and the attacker's guessed value.

\begin{figure}[h]
	\vspace{-1mm}
	\begin{center}
            \includegraphics[width=0.30\textwidth]{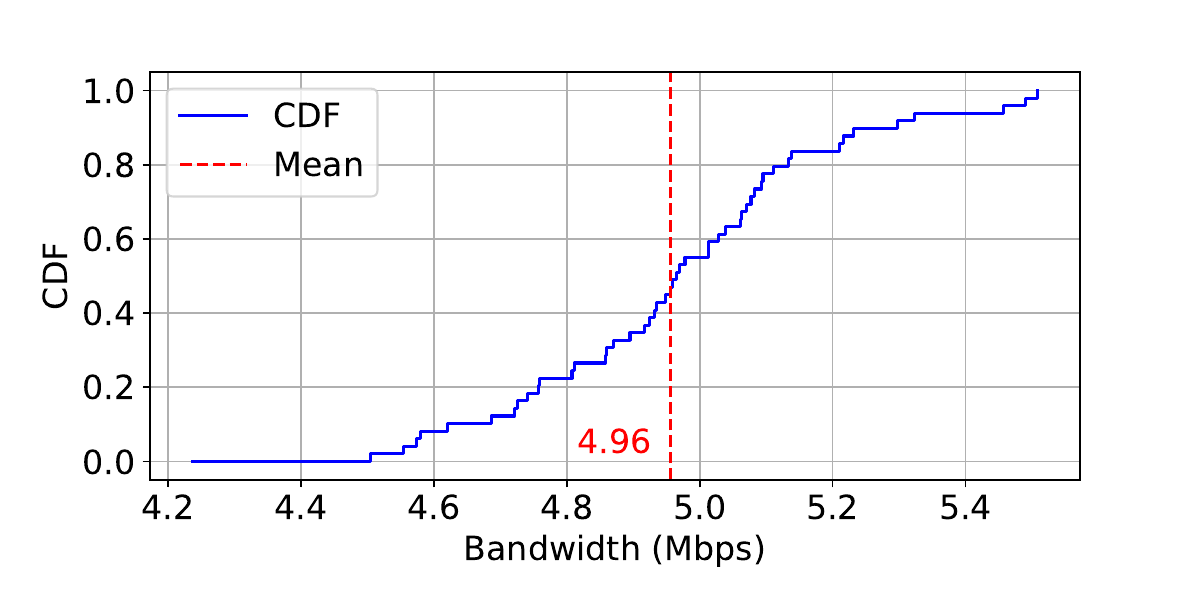}
		\vspace{-2mm}
		\caption{Attack traffic used to send crafted TCP \texttt{RST} packets to achieve a DoS attack success rate of 70\% on average.}
		\label{pic:rstband}
	\end{center}
	\vspace{-4mm}
\end{figure}

In practice, the attacker can send multiple TCP \texttt{RST} packets in parallel, covering a range of sequence numbers near the inferred sequence number, thereby increasing the attack's success rate.
Figure~\ref{pic:rstband} illustrates the attack traffic bandwidth used to send TCP \texttt{RST} packets.
In summary, considering the time required to identify the victim TCP connection (i.e., 3.80 seconds for the sequential allocation method or 21.82 seconds for the preserved allocation method) and to infer the sequence numbers of the victim TCP connection (194.21 seconds), as well as the maximum attack traffic bandwidth needed for the attack (either 11.63 Mbps for inferring the sequence numbers or 4.96 Mbps for sending crafted \texttt{RST} packets), the DoS attack can be successfully performed with 211 seconds on average, achieving a success rate of 70\%.

\subsubsection{Poisoning FTP}
The attacker injects crafted TCP data packets into the identified FTP session. These forged data packets carry an inferred acceptable sequence number that falls within the victim device's receive window (i.e., within the range of [$rcv.nxt$, $rcv.nxt$+$rcv.wnd$]), along with a brute-forced acknowledgment number.
Once the acknowledgment number specified in the packet falls within the victim device's acceptable acknowledgment number range, the victim accepts and processes the crafted TCP packet, ultimately poisoning the received FTP file.

\begin{figure}[h]
	\vspace{-2mm}
	\begin{center}
            \includegraphics[width=0.4\textwidth]{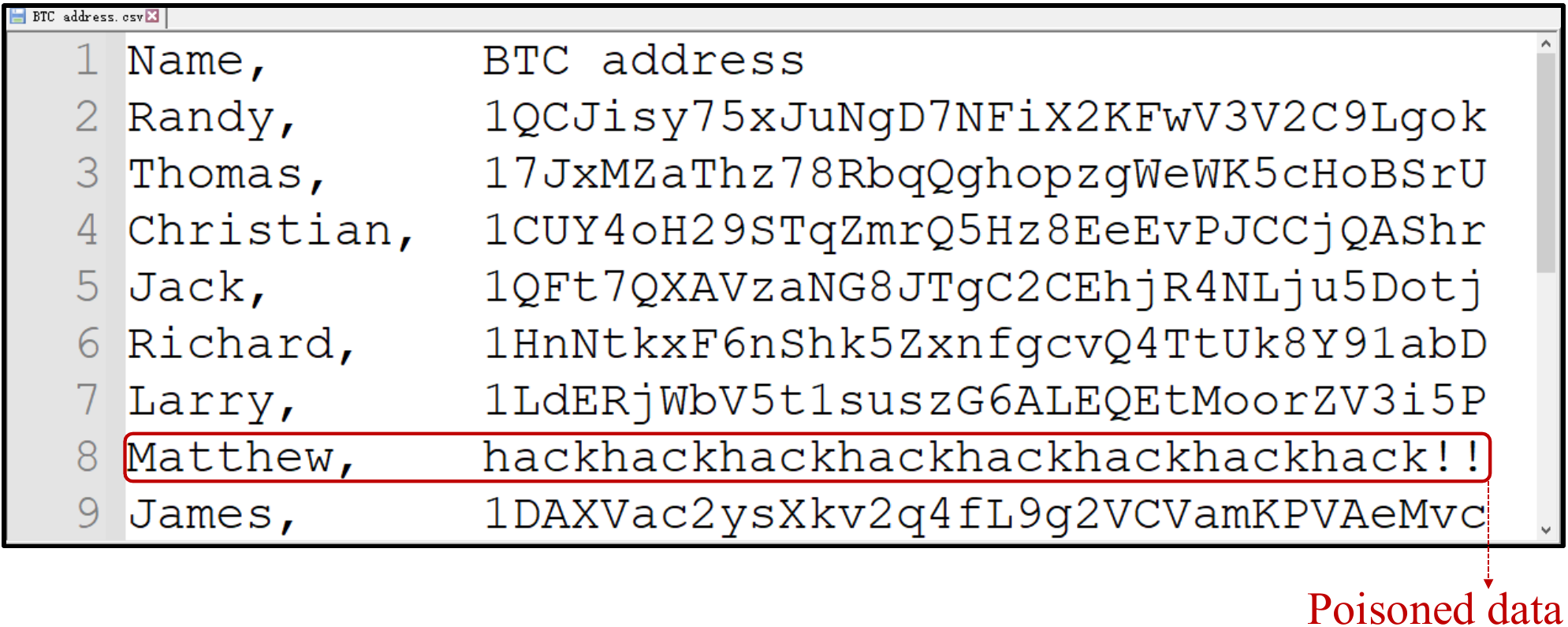}
		\vspace{-2mm}
		\caption{The poisoned FTP file received by the victim device.}
		\label{pic:ftp-poisoned}
	\end{center}
	\vspace{-2mm}
\end{figure}

\begin{figure}[h]
    \vspace{-4mm}
	\begin{center}
		\subfigure[Traffic cost for Linux.]{ 
			\label{pic:port_time}  
			\includegraphics[width=0.22\textwidth]{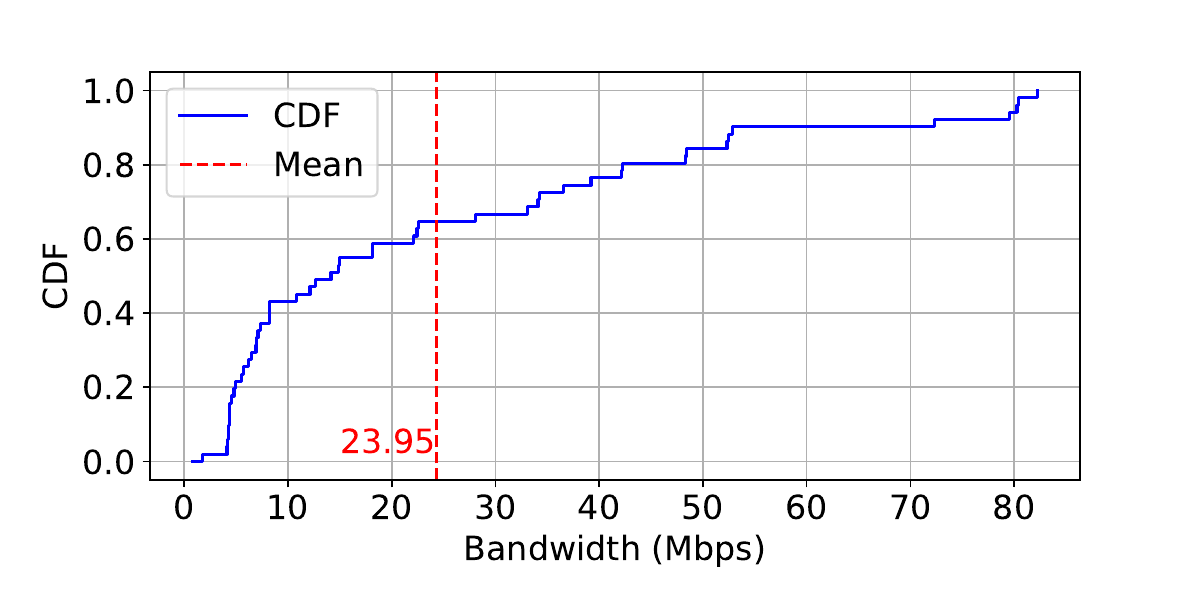} 
		} 
        \subfigure[Traffic cost for OpenBSD.]{ 
			\label{pic:port_time}  
			\includegraphics[width=0.22\textwidth]{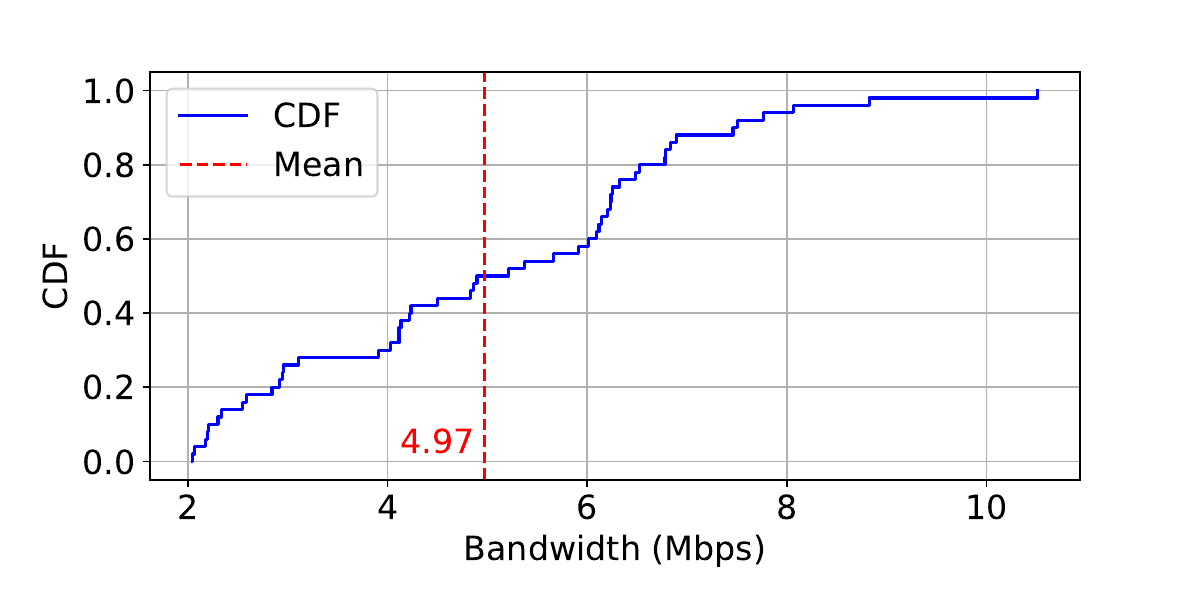} 
		} 
		\subfigure[Traffic cost for Windows.]{ 
			\label{pic:port_time}  
			\includegraphics[width=0.22\textwidth]{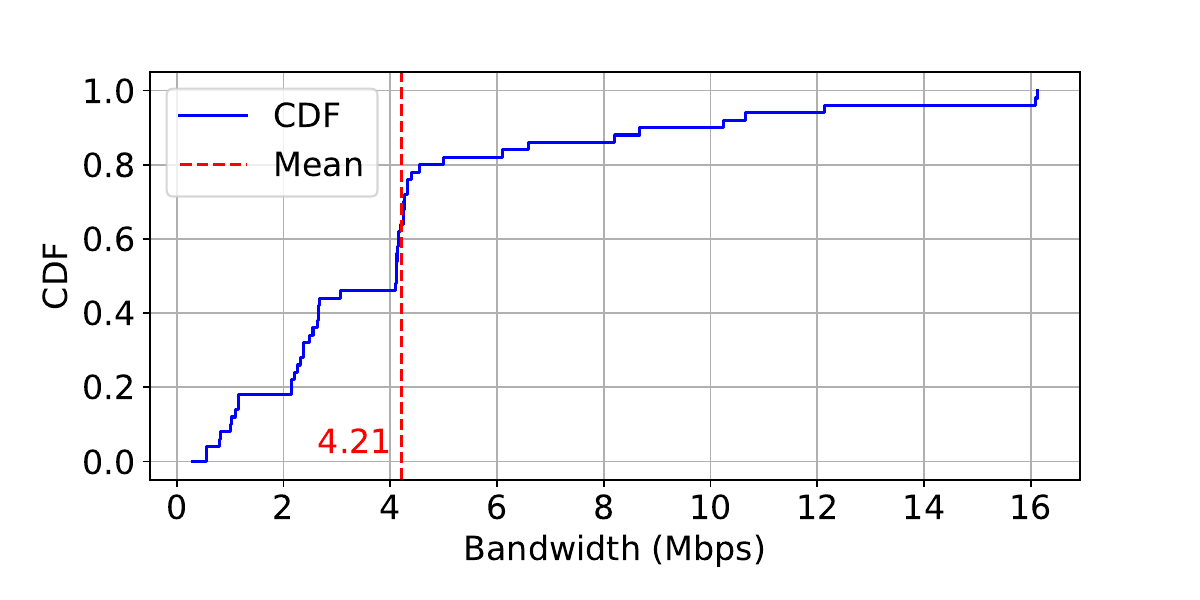} 
		} 
        \subfigure[Traffic cost for macOS.]{ 
			\label{pic:port_time}  
			\includegraphics[width=0.22\textwidth]{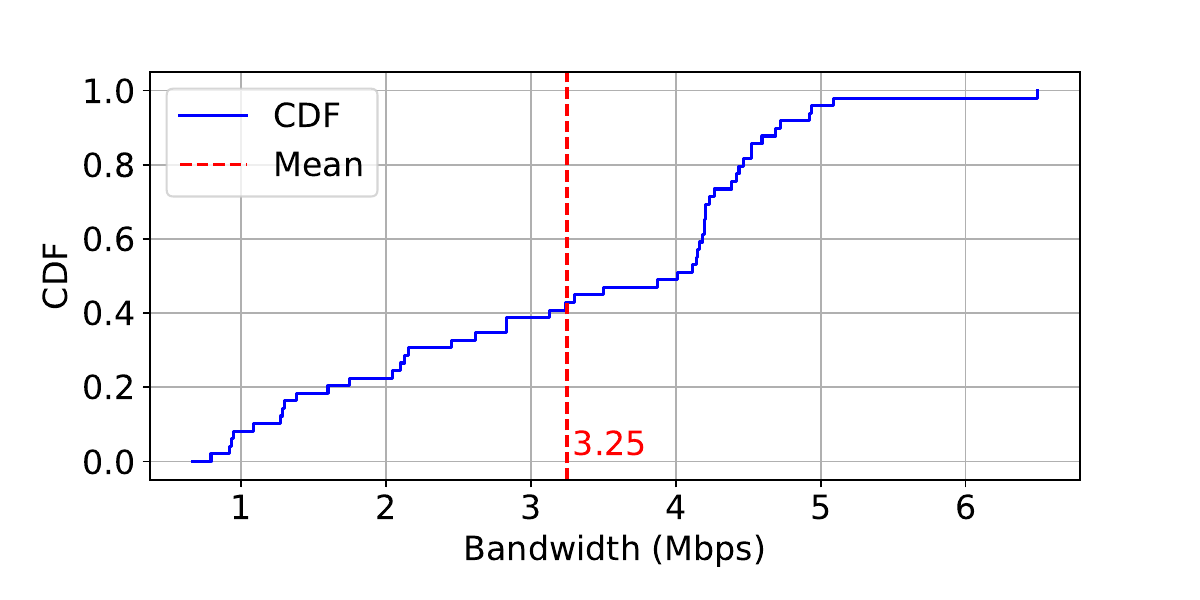} 
		} 
        \vspace{-2mm}
		\caption{CDF of attack traffic costs for poisoning the received FTP file under different victim device OSes.} 
		\label{pic:posing-brute} 
	\end{center}
	\vspace{-3mm}
\end{figure}

Figure~\ref{pic:ftp-poisoned} shows the snapshot of the poisoned FTP file received by the victim device.
Figure~\ref{pic:posing-brute} illustrates the attack traffic bandwidth used to brute-force the acknowledgment numbers, thus successfully achieving FTP file poisoning against the victim device across different OSes.
For instance, on Linux 5.15 system, the average attack bandwidth for FTP poisoning is 23.95 Mbps, whereas for OpenBSD 7.3, Windows 11 Pro 23H2, and macOS 14.2, the needed bandwidth is 4.97 Mbps, 4.21 Mbps, and 3.25 Mbps, respectively.
We conduct 50 experiments, and with the attack bandwidth shown in Figure~\ref{pic:posing-brute}.

In total, accounting for the time required to identify the victim TCP connection and infer its sequence numbers, as well as the maximum attack traffic bandwidth needed (i.e., the greater of the bandwidth for inferring sequence numbers or for brute-forcing the acknowledgment number across different OSes), the FTP poisoning attack can be successfully executed with 217 seconds on average using an attack traffic bandwidth of 23.95 Mbps\footnote{The bandwidth of 23.95 Mbps required for brute-forcing the acknowledgment number during an attack on a Linux system represents the maximum needed for conducting an FTP poisoning attack. For other OSes, the required bandwidth is 11.63 Mbps for inferring the victim TCP connection's sequence numbers, as brute-forcing their acknowledgment numbers requires less bandwidth.}, achieving a success rate of 72\% (36 out of 50).

\subsubsection{HTTP Injection}
\begin{figure}[h]
	\vspace{-2mm}
	\begin{center}
            \includegraphics[width=0.42\textwidth, height=0.25\textwidth]{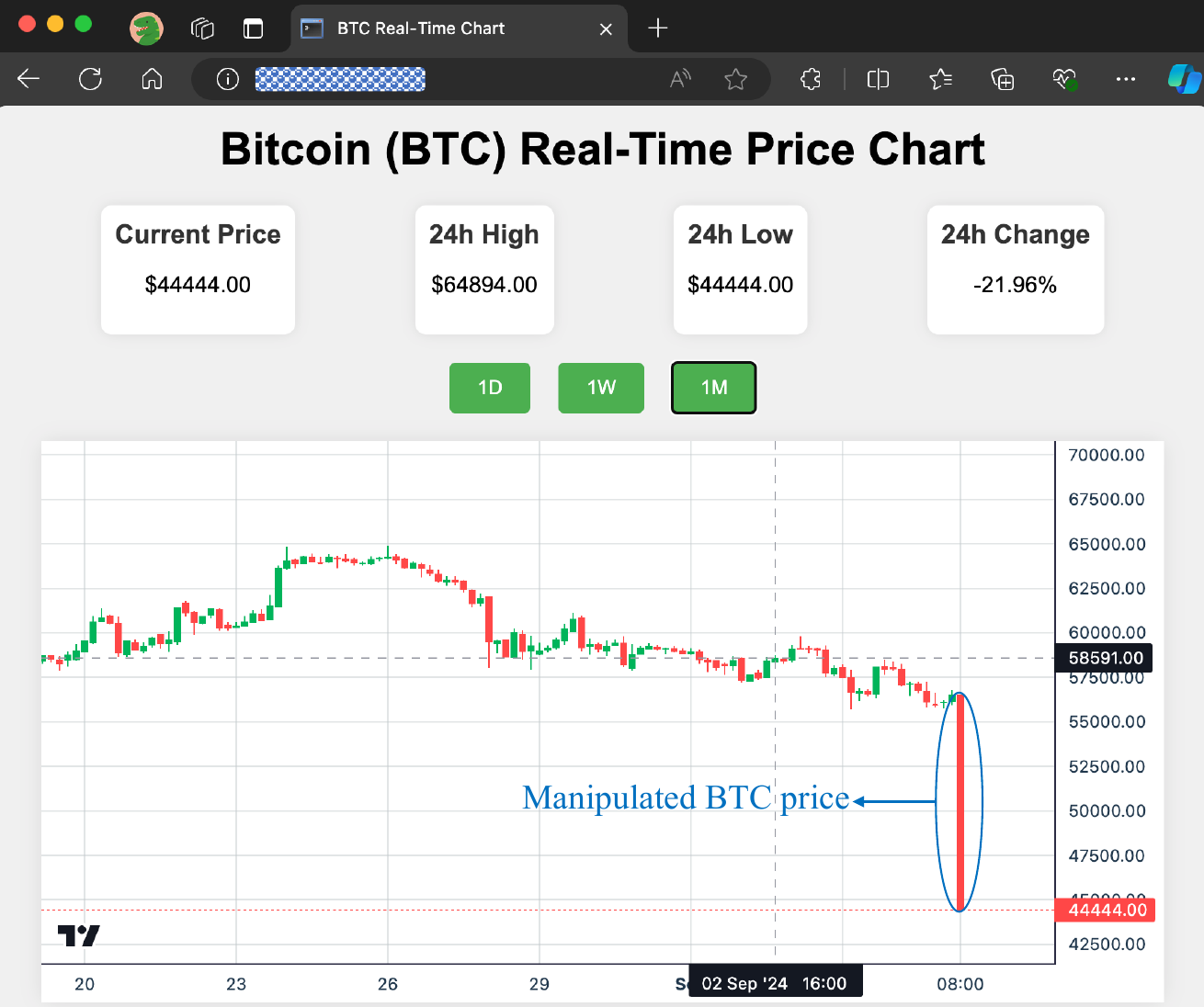}
		\vspace{-2mm}
		\caption{Poisoned HTTP message received by the victim.}
		\label{pic:http-poisoned}
	\end{center}
	\vspace{-4mm}
\end{figure}

Similarly, the attacker can craft TCP data packets to poison HTTP. For instance, the attacker can detect a TCP connection from a specific source port of the gateway to the server's port 80, then inject a fake HTTP message into this connection, thereby poisoning the victim device’s browser content. We evaluate this attack by injecting fake HTTP messages into a Windows victim device. In 50 experiments, the results show that with the time cost of 212 seconds on average, using approximately 11.63 Mbps of bandwidth, the attacker successfully injects fake HTTP messages into the Windows device in 37 out of 50 attempts, achieving a success rate of 74\%.
Figure~\ref{pic:http-poisoned} shows a snapshot of the poisoned HTTP message received by the victim device, where the attacker manipulated the Bitcoin price.

\section{Real-World Measurements}
\label{sec:real-world}

In this section, we measure 50 real-world networks with IP address sharing enabled, covering a diverse range of scenarios, including public Wi-Fi networks in coffee shops, hotels, and other locations, as well as 5G cellular and VPN networks. Our study reveals that 38 out of the 50 tested networks (76\%) are vulnerable.

\subsection{Experimental Setup \& Workflow}
In our real-world measurement study on the Internet, we deploy two devices: a victim device (Samsung Galaxy S22) and a puppet (Ubuntu 22.04 with Linux kernel 5.15) within target networks that enable IP address sharing.
For public Wi-Fi networks with NAT enabled, both devices connect to the NAT gateway (typically the wireless router holding the shared public IP address), following approval from the network manager.
For 5G networks, we deploy a Customer Premises Equipment (CPE) device within the target network, with both the victim device and the puppet connected to this CPE. This ensures our experiments are fully controlled and do not affect regular mobile users (i.e., posing no ethical issues). The NAT-enabled User Plane Function (UPF) in the 5G core network (5GC) assigns a shared public IP address and ephemeral source ports for TCP connections initiated by the victim device and the puppet.
In the VPN scenario, the victim device and the puppet connect to a VPN proxy server (e.g., WireGuard deployed on the Internet under our control). The VPN proxy server assigns a shared public IP address and associated ephemeral source ports for their TCP connections.
%
The victim device and the puppet establish their respective TCP connections with a remote server, a VPS we control running Linux 5.15 in California, USA.
%
%
An attacker on the Internet, capable of source IP address spoofing and running Kali 2022.2, attempts to hijack the TCP connection between the victim device and the server. The attacker may launch either a DoS attack against SSH or a poisoning attack against FTP traffic. If these attacks can be successfully executed, we consider the target network as vulnerable.

\begin{table*}[h]
\centering
\small
\renewcommand\arraystretch{0.95}
\caption{Evaluations of off-path TCP connection hijacking on 50 real-world networks enabling IP address sharing.}
\vspace{-2mm}
\label{tab:real-world-evaluations}
\resizebox{0.95\linewidth}{!}{%
\begin{tabular}{ccclrccccc}
\bottomrule
\multirow{2}{*}{\textbf{No.}} & \multirow{2}{*}{\textbf{\begin{tabular}[c]{@{}c@{}}The Shared IP Address of\\ the Accessed Network\end{tabular}}} & \multirow{2}{*}{\textbf{CIDR}} & \multirow{2}{*}{\textbf{Network Access Method}} & \multirow{2}{*}{\textbf{\begin{tabular}[c]{@{}c@{}}Port\\ Assignment\end{tabular}}} & \multirow{2}{*}{\textbf{Vulnerable}} & \multicolumn{2}{c}{\textbf{Time Cost}} & \multicolumn{2}{c}{\textbf{Success Rate}} \\
 &  &  &  &  &  & \textbf{SSH DoS (s)} & \textbf{FTP Poisoning (s)} & \textbf{SSH DoS} & \textbf{FTP Poisoning} \\ \midrule

1 & ***.***.22.43  &/24& Wi-Fi at coffee shop & preserved & \blackcheck & 243.72 & 202.53 & 6/10 & 5/10 \\
2 & ***.***.113.13  &/18& Wi-Fi at coffee shop & preserved & \blackcheck & 180.08& 166.19& 8/10& 6/10\\
3 &  ***.***.10.160& /18 & Wi-Fi at hotel & preserved & \blackcheck & 197.15 & 208.78 & 7/10 & 6/10 \\ 
4 &  ***.***.188.126 & /18 & Wi-Fi at hotel& preserved & \blackcheck & 169.37 & 203.32 & 8/10 & 6/10 \\ 
5 & ***.***.185.110 & /18 & Wi-Fi at restaurants  & preserved & \blackcheck & 176.16 & 194.21 &6/10&7/10 \\ 
6 & ***.***.131.17 & /24 & Wi-Fi at restaurants & preserved & \blackcheck & 217.30 & 179.62 & 6/10 & 6/10 \\ 
7 & ***.***.230.45 & /18 & Wi-Fi at restaurants& preserved & \blackcheck & 186.70 & 197.36 & 7/10 & 6/10 \\  
8 & ***.***.87.142 & /18 & Wi-Fi at Shopping Mall & preserved & \blackcheck & 183.29 & 206.16 & 8/10 & 6/10 \\ 
9 & ***.***.121.148 & /18 & Wi-Fi at Shopping Mall & preserved & \blackcheck & 208.26 & 185.90 & 7/10 & 8/10 \\ 
10 & ***.***.239.211 & /18 & Wi-Fi at campus & sequential & \blackcheck & 191.64 & 195.25 & 8/10 & 5/10 \\ 
11 & ***.***.68.231 & /18 & Wi-Fi at campus & sequential & \blackcheck & 197.64 & 207.63 & 8/10 & 6/10 \\
12 & ***.***.118.14 & /23 & Wi-Fi at Company & sequential & \blackcheck & 200.67 & 212.13 & 7/10 & 7/10 \\

\rowcolor{mygray}13 & ***.***.105.29 & /18 & Wi-Fi at campus & random & \blackcross & N/A & N/A & N/A & N/A \\
\rowcolor{mygray}14 & ***.***.46.163&/18& Wi-Fi at coffee shop & random & \blackcross & N/A & N/A & N/A & N/A \\ 
\rowcolor{mygray}15 &  ***.***.115.41& /18 & Wi-Fi at hotel & random & \blackcross & N/A & N/A & N/A & N/A \\ 
\rowcolor{mygray}16 & ***.***.120.129  & /18 & Wi-Fi at hotel & random & \blackcross & N/A & N/A & N/A & N/A \\
\rowcolor{mygray}17 & ***.***.186.95 &/18  & Wi-Fi at restaurants  & random & \blackcross & N/A & N/A & N/A & N/A \\ 
\rowcolor{mygray}18 &  ***.***.15.200& /22 & Wi-Fi at restaurants & random & \blackcross & N/A & N/A & N/A & N/A \\ 
\rowcolor{mygray}19 & ***.***.35.107 & /18 & Wi-Fi at Shopping Mall & random & \blackcross & N/A & N/A & N/A & N/A \\ 
\rowcolor{mygray}20 & ***.***.64.96 & /22  &  Wi-Fi at Shopping Mall & random & \blackcross & N/A & N/A & N/A & N/A \\

\midrule
21 & ***.***.40.67 & /22 & 5G at Airport & sequential & \blackcheck & 223.89 & 211.34 & 8/10 & 7/10 \\
22 & ***.***.41.73 & /22 & 5G at Airport & sequential & \blackcheck & 211.79 & 197.75 & 8/10 & 8/10 \\
23 & ***.***.237.90 & /18 & 5G at Airport & sequential & \blackcheck & 182.81 & 230.14 & 9/10 & 8/10 \\
24 & ***.***.239.175 & /18 & 5G at Airport & sequential & \blackcheck & 203.93 & 184.36 & 7/10 & 8/10 \\
25 & ***.***.91.199 & /17 & 5G at Airport & sequential & \blackcheck & 176.39 & 198.50 & 7/10 & 7/10 \\
26 & ***.***.1.170 & /24 & 5G at Airport & sequential & \blackcheck & 210.69 & 206.04 & 10/10 & 4/10 \\
27 & ***.***.237.118 & /18 & 5G at Train Station & sequential & \blackcheck & 212.58 & 314.99 & 8/10 & 7/10 \\
28 & ***.***.40.218 & /22 & 5G at Train Station & sequential & \blackcheck & 210.85 & 199.27 & 7/10 & 8/10 \\
29 & ***.***.0.247 & /24 & 5G at Train Station & sequential & \blackcheck & 176.98 & 198.71 & 8/10 & 7/10 \\
30 & ***.***.41.207 & /22 & 5G at Train Station & sequential & \blackcheck & 200.93 & 218.27 & 8/10 & 6/10 \\
31 & ***.***.237.190 & /18 & 5G at Train Station & sequential & \blackcheck & 206.91 & 203.20 & 7/10 & 5/10 \\
32 & ***.***.2.74 & /24 & 5G at Office Building & sequential & \blackcheck & 265.58 & 266.09 & 9/10 & 7/10 \\
33 & ***.***.116.83 & /24 & 5G at Office Building & sequential & \blackcheck & 162.13 & 192.67 & 9/10 & 7/10 \\
34 & ***.***.2.35 & /24 & 5G at Office Building & sequential & \blackcheck & 205.98 & 256.79 & 10/10 & 4/10 \\
35 & ***.***.22.220 & /24 & 5G at Office Building & sequential & \blackcheck & 140.73 & 289.70 & 7/10 & 6/10 \\
36 & ***.***.237.202 & /18 & 5G at Office Building & sequential & \blackcheck & 220.53 & 190.50 & 8/10 & 9/10 \\
37 & ***.***.236.121 & /18 & 5G at Office Building & sequential & \blackcheck & 197.67 & 286.00 & 8/10 & 7/10 \\
38 & ***.***.40.171 & /22 & 5G at Office Building & sequential & \blackcheck & 186.75 & 228.64 & 6/10 & 4/10 \\
39 & ***.***.41.177 & /22 & 5G at Campus & sequential & \blackcheck & 162.47 & 245.61 & 7/10 & 8/10 \\
40 & ***.***.238.188 & /18 & 5G at Campus & sequential & \blackcheck & 213.58 & 264.87 & 7/10 & 7/10 \\
\midrule

41 & ***.***.239.144 & /15 & VPN proxy of WireGuard & preserved & \blackcheck & 172.35 & 285.48 & 7/10 & 5/10 \\
42 & ***.***.41.227 & /18 & VPN proxy of PPTP & preserved & \blackcheck & 256.89 & 241.59 & 9/10 & 9/10 \\
43 & ***.***.236.119 & /18 & VPN proxy of L2TP & preserved & \blackcheck & 183.84 & 283.66 & 7/10 & 9/10  \\
44 & ***.***.3.228 & /15 & VPN proxy of OpenVPN 2.6.9 & preserved & \blackcheck & 248.13 & 213.94 & 10/10 & 7/10 \\

45 & ***.***.0.245 & /15 & VPN proxy of OpenConnect 1.5.3 & preserved & \blackcheck & 258.83 & 292.06 & 4/10  & 7/10 \\
46 & ***.***.237.68 & /18 & VPN proxy of OpenConnect 1.6.2 & preserved & \blackcheck & 247.82 & 245.63 & 7/10 & 8/10 \\
\rowcolor{mygray}47 & ***.***.40.122 & /15 & VPN proxy of Trojan & random & \blackcross & N/A & N/A & N/A & N/A \\ 
\rowcolor{mygray}48 & ***.***.40.122 & /15 & VPN proxy of ShadowSocks & random & \blackcross & N/A & N/A & N/A & N/A \\ 
\rowcolor{mygray}49 & ***.***.2.166 & /15 & VPN proxy of Vmess & random & \blackcross & N/A & N/A & N/A & N/A \\
\rowcolor{mygray} 50 & ***.***.116.180 & /24 & VPN proxy of EtherVPN & random & \blackcross & N/A & N/A & N/A & N/A \\

\toprule
\end{tabular}
}
\vspace{-0.8mm}
\end{table*}

\subsection{Experimental Results}
Table~\ref{tab:real-world-evaluations} presents the experimental results of our attacks on 50 real-world networks enabling IP address sharing. Overall, 38 of the 50 tested networks are found to be vulnerable, while the 12 invulnerable networks are protected by the \textit{Random Allocation} strategy adopted by the gateway to assign ephemeral TCP source ports to connected devices.
As depicted in the first row of the table, we discover that a NAT-enabled public Wi-Fi network with a shared public IP address of ``***.***.22.43''\footnote{For ethical reasons, we anonymize the first two segments of the public IP address for the accessed networks.} can be accessed from a coffee shop. It has a CIDR of /24. The gateway of this Wi-Fi network employs the preserved method to assign ephemeral source ports for outgoing TCP connection requests, making it vulnerable to our attack.
The time required to execute our attack to compromise TCP connections initiated from this network is 243.72 seconds to terminate the SSH connection and 202.53 seconds to poison the FTP files on average. The success rates is 6 out of 10 for terminating the SSH connection and 5 out of 10 for FTP poisoning.
Attack failures can be attributed to network noise and potential interception of crafted packets by firewalls deployed within the network infrastructure.

Regarding the 20 NAT-enabled 5G networks with public IP address sharing we tested, all of them use a sequential-based method (i.e., the \textit{Per-Destination Sequential Allocation}) to assign ephemeral source ports for outgoing TCP connection requests. As a result, all 30 tested 5G networks are vulnerable to our attack. Specifically, 5G networks typically have a large number of concurrent users, and the sequential-based method effectively avoids port collisions, which is crucial for both network performance and user experience.
Moreover, our attack can also compromise VPNs. Testing 10 VPN proxy servers with shared public IPs, we find 6 vulnerable due to flawed ephemeral port allocation. 
%
These findings demonstrate the severity of our attack on Internet security.





%


\section{Discussion and Countermeasure}
\label{sec:countermeasure}

\subsection{Attack Limitations}

\noindent\textbf{i) ICMP Error Messages Blocking.}
In our attack, the attacker needs to send crafted ICMP ``Packet Too Big'' messages to the server to infer the sequence numbers of the identified TCP connection.
%
In practice, firewalls on real-world servers may block the received ICMP ``Packet Too Big'' messages, resulting in failed sequence number inference.
We ethically test whether web servers in the Tranco Top 10K sites list accept ICMP ``Packet Too Big'' messages. We establish a TCP connection between our controlled client and the server, and then send an ICMP ``Packet Too Big'' message to the server.
If the server reduces the TCP packet size for our client accordingly, it indicates that the server does not block the ICMP message.
The experimental findings reveal that among the Tranco Top 10K sites, 471 sites (4.71\%) block ICMP ``Packet Too Big'' messages to prevent our attack, even though the blocking of the ICMP messages potentially introduces connection failures in practice. 

\noindent\textbf{ii) Specific to Newly Established TCP Connections.}
%
Another limitation of our attack is its specificity to newly established TCP connections initiated by the victim device. As shown in Figure~\ref{pic:Linuxport} and Figure~\ref{pic:portidentify}, our attack relies on occupying a continuous range of ephemeral source ports on the gateway or identifying the baseline of sequential port allocation first. Then, the attacker waits for the victim  client to initiate a new TCP connection to the target server, allowing the identification of the ephemeral source port allocated by the gateway for this newly established TCP connection.
This identification can be achieved either by measuring the time delay caused by port collisions
or by testing the occupation of the subsequent port next to the baseline of sequential port allocation.
In practice, attackers can periodically probe the gateway after waiting for a specified time window to determine whether an ephemeral source port for the victim server is occupied, enabling them to identify a victim TCP connection. This approach mitigates the limitation.

\noindent\textbf{iii) Attack Impact on IPv6 Networks.}
IPv6 mitigates our attack since IP address sharing is less common in IPv6 networks. However, NAT46 and even IPv6-only environments (e.g., VPN proxies where multiple IPv6 clients share the same IPv6 address) are still vulnerable to our attack. We test that IPv6 servers' TCP sequence numbers can be inferred via the PMTUD side channel when IPv6 address sharing is enabled.

\subsection{Countermeasure}

%
PMTUD operates on IP address pairs, thus making it impossible to achieve information isolation between different TCP connections to the same IP address (e.g., to a NAT network or to a VPN proxy server that uses a shared public IP address for multiple devices).
Therefore, off-path attackers on the Internet can forge ICMP ``Packet Too Big'' messages to infer the sequence numbers of victim TCP connections.
We propose that PMTUD should operate on a per-connection basis rather than a per-IP basis. This approach would effectively prevent information leakage between different TCP connections sharing the same IP address in various real-world network scenarios. In practice, IP packets belonging to different connections may traverse different network paths to the same destination. Therefore, the path MTU of one connection should not dictate the path MTU for all connections to the same destination IP. We have reported this vulnerability and our countermeasure to the IETF.
Following the IETF's suggestions, we will present our findings at a working group meeting and propose standardization to address this vulnerability.

Additionally, ISPs can adopt a randomized method for assigning ephemeral source ports to TCP connections at gateways with IP address sharing, instead of the \textit{Port Preservation} or \textit{Per-Destination Sequential Allocation} methods. However, randomization may increase port collisions, which could reduce network performance, particularly in environments with many concurrent users.
Nevertheless, this approach prevents off-path attackers from predicting the distribution of ephemeral source ports of TCP connections originating from the same gateway, thereby thwarting the attacker’s ability to identify TCP connections.
%
%
We have disclosed the vulnerable source port allocation methods to the identified vendors—HUAWEI, H3C, Cisco (through their Bug Bounty Program), and the Linux community—which have been recognized by Linux, Cisco, and H3C.
In particular, Cisco has taken steps to mitigate the vulnerability in its products, for example, by introducing randomized ephemeral source port allocation in firmware versions 16.12.85 and 17.03.08.




\section{Related Work}
\label{sec:related}


\noindent\textbf{Off-Path TCP Exploits.}
Off-path TCP exploits pose a significant threat to Internet security.
%
By exploiting the global IPID (IP Identification) counter used by earlier OSes, Gilad \etal were able to determine whether two hosts had established a TCP connection. Subsequently, they initiated off-path TCP injection attacks aimed at poisoning the upper HTTP or Tor traffic~\cite{gilad2012spying,gilad2013off,gilad2012off}.
Feng \etal demonstrated the vulnerability of mixed IPID assignment in modern Linux systems, which can be exploited by off-path attackers to hijack TCP connections~\cite{ccsfeng}. This vulnerability persists even under the IPv4/IPv6 dual-stack~\cite{feng2021off}.
Cao \etal demonstrated that an off-path attacker could exploit a side channel in the challenge ACK mechanism to hijack TCP connections, which was mitigated by a random challenge ACK count limit~\cite{cao2018off,cao2016off}. A timing side channel in Wi-Fi was also identified, enabling data injection into a TCP connection to cache malicious objects~\cite{chen2018off}.
%
%
%
Yang \etal discovered that session mappings in certain NAT implementations are vulnerable to manipulation by fake TCP \texttt{RST} packets, allowing attackers to intercept a victim client's TCP packets and hijack the connection by stealing sequence numbers~\cite{yangexploiting}. 
%
However, this attack relies on vulnerabilities in specific NAT implementations, which have been patched in versions like OpenWrt 23.05 and FreeBSD 14.0 and later.

Attackers can also exploit unprivileged scripts
running on compromised hosts to carry out off-path TCP attacks~\cite{qian2012collaborative, qian2012off, gilad2013tolerance}.
Gilad \etal demonstrated that attackers can execute web cache poisoning by exploiting restricted scripts within the sandbox of a user's browser~\cite{gilad2013tolerance}.
Qian \etal revealed that firewalls, acting as middleboxes, can be exploited to infer TCP sequence numbers based on how they differently handle packets with in-window and out-of-window sequence numbers~\cite{qian2012off}
Additionally, assisted by malicious applications, they conducted a collaborative attack on TCP sequence number inference by exploiting side channels related to packet counters~\cite{qian2012collaborative}.
Tolley \etal demonstrated that on-path/in-path attackers in VPN scenarios can infer TCP sequence numbers~\cite{tolley2021blind}.


\noindent\textbf{Port De-Randomization.}
De-randomizing the UDP source port of DNS requests has been well studied in prior works~\cite{man2020dns, man2021dns, Herzberg2012SecurityOP,cross2008dns,alharbi2019collaborative}. For example, attackers may identify a UDP source port for DNS using a ``reserve-and-trap'' method, where a malicious NATed device issues multiple UDP sessions to occupy the NAT gateway’s source ports (e.g., up to 65,534 ports, leaving only one predictable port available)~\cite{Herzberg2012SecurityOP,alharbi2019collaborative}. However, we identify that this method fails on modern NAT implementations (e.g., Linux 5.0 and later) due to multiple source port collisions, preventing the victim device from acquiring the intended port. For NAT gateways using Linux 5.0 and later, if port collisions exceed 240, ephemeral port allocation fails.
An attacker can use malicious JavaScript to de-randomize the Simple Hash-Based Port Selection (SHPS) algorithm in older Linux versions and identify a TCP source port to compromise the connection~\cite{gilad2014off}.
Our method can identify the ephemeral source ports of gateways, even when the victim device uses the newer Double-Hash Port Selection (DHPS) algorithm—without relying on JavaScript or spoofing.



\noindent\textbf{PMTUD Issues.}
PMTUD is designed to protect TCP from IP fragmentation~\cite{herzberg2013fragmentation, gilad2013fragmentation, gilad2012off}.
G{\"o}hring \etal showed that attackers could exploit PMTUD to induce performance degradation in TCP~\cite{gohring2018path}.
RFC 6269 highlights IP address sharing issues, including path MTU cache inconsistencies and DoS risks from a malicious subscriber (i.e., the puppet in our context) reducing the MTU below 68 octets~\cite{rfc6269}.

Feng \etal demonstrated that an off-path attacker can bypass PMTUD due to ambiguities in the cross-layer interactions among IP, ICMP, and TCP. This allows the attacker to trigger IP fragmentation on TCP and poison the target TCP traffic with malicious IP fragments~\cite{feng2022ndss}.
They also demonstrated that a vantage point (i.e., a malicious server) on the Internet can determine whether a client accessing it is a NATed host by issuing an ICMP message containing known information about the connection between the server and the client~\cite{feng2025ndss}. This message induces a desynchronization of the path MTU value between the NAT gateway and the NATed host.
Distinctively, our attack leverages a new side channel to infer unknown TCP sequence numbers in the off-path threat model.
Our attack evades detection by existing techniques~\cite{cao2019principled,ru2021side}.

\section{Conclusion}
\label{sec:conclusion}

In this paper, we investigate the security implications arising from the interactions between PMTUD and IP address sharing, two fundamental components of Internet infrastructure. We uncover that PMTUD is not well-suited for networks with IP address sharing, leading to vulnerabilities that off-path attackers can exploit to infer sequence numbers of victim TCP connections.
The root cause is PMTUD's inability to maintain per-connection information isolation, enabling attackers to send forged ICMP ``Packet Too Big'' messages and observe changes in the TCP packet sizes sent by the server.
Through extensive experiments, we demonstrate that this vulnerability can be exploited to orchestrate off-path TCP hijacking attacks, compromising various applications by maliciously terminating SSH connections or poisoning FTP and HTTP traffic, which can cause significant real-world damage. To address this attack, we propose effective countermeasures.


\begin{acks}
We thank the anonymous reviewers for their insightful comments. This work was supported in part by the National Key Research and Development Program of China under Grant 2022YFB3102303, the National Science Foundation for Distinguished Young Scholars of China under No. 62425201, the Science Fund for Creative Research Groups of the National Natural Science Foundation of China under No. 62221003, the Key Program of the National Natural Science Foundation of China under No. 61932016 and No. 62132011, and the State Key Laboratory of Internet Architecture.

\end{acks}

\bibliographystyle{ACM-Reference-Format}
\balance
\bibliography{reference}



\end{document}